\documentclass[twocolumn,showpacs,preprintnumbers,amsmath,amssymb,prb,superscriptaddress]{revtex4}

\usepackage{graphicx}

\begin{document}
 
\title{Analytical results for Green's functions of lattice fermions}
\author{A. Komnik}
\author{S. Heinze}
\affiliation{Institut f\"ur Theoretische Physik, Universit\"at Heidelberg, Philosophenweg 12, D-69120 Heidelberg, Germany}

\date{\today}

\begin{abstract}

We present a further development of methods for analytical calculations of Green's functions of lattice fermions based on recurrence relations. Applying it to tight-binding systems and topological superconductors in different dimensions we obtain a number of new results. In particular we derive an explicit expression for arbitrary Green's function of an open Kitaev chain and discover non-local fermionic corner states in a 2D $p$-wave superconductor.  
\end{abstract}


\maketitle

\section{Introduction}
\label{sec:intro}

Fermionic lattice models are widely used not only as a purely theoretical tool but also as a basis for investigation and modelling of physical properties of real materials. Despite their relative formal simplicity -- the Hamiltonians of many of them can be written down as bilinears of fermionic operators -- many of them resist explicit analytical solution. The most prominent example is the 
Azbel-Hofstadter problem of lattice fermions subject to a magnetic field. \cite{Azbel64,Hofstadter76} Although it is possible to set up recurrence relations for relevant observables, an explicit analytical solution in a closed form has not yet been found. Remarkably, the presence of the gauge field complicates the progress making necessary the application of such advanced methods as Bethe ansatz.\cite{WIEGMANN1994495,Wiegmann94,ABANOV1998571} 
The situation is better for special constellations of fields, when some analytical solutions are possible, see e.~g. [\onlinecite{Hatsugai93PRB,Ueta1997}].
But also in simpler field free situations analytical results for solutions of recurrence relations are rare and far between, vast majority of studies being concentrated on numerical treatment of the problem.\cite{LeeJoannopoulos1981,SurfaceToBulk2011,VelevButler_2004}

If one restricts oneself to a system's single particle Green's functions (GFs) an explicit solution on the level of eigenstates is not always necessary. Here a formal functional integral for GFs can immediately be written down. However, its computation turns out to be very cumbersome. This kind of a functional integral is essentially a sum over all possible paths the particle can take during its evolution between two states. While in the absence of the gauge field the phase gathered along each individual path is simply related to the path length, in presence of the field it acquires a highly non-trivial dependence on the path geometry and topology. One attempt to take that into account is presented in [\onlinecite{Belov1991}]. However, the resulting expressions are complicated and very difficult to handle.

Recurrence relations for GFs were originally proposed in the series of works,\cite{Haydock1,Haydock_recursion} and subsequently successfully used for numerical calculations in a great variety of set-ups, see e.~g. [\onlinecite{MacKinnon:1985rt}]. In recent years this method became very popular in the field of topological insulators as it allows for fast and efficient band structure calculations in systems with any kinds of elementary cells and arbitrary coupling mechanisms between them, see e.~g. [\onlinecite{Chinese2006,KlichLong2014,DwivediChua16}]. On the other hand, in many situations the recurrence relations can be solved analytically leading to compact and useful results for relevant physical quantities.\cite{DoornenbalStoof2015,MajoranaAK2016} The goal of our paper is twofold: first we apply this efficient technique to systems in gauge fields and second, we present new analytical results not only for systems with topologically non-trivial band structures but also for simple cubic lattices in different dimensions for different surface and bulk geometries.

The structure of our presentation is as follows: In Section \ref{Basics} we explain our ideas on the simple example of a spinless 1D tight-binding chain. Among other things we present  a new analytical result Eq.~(\ref{throughU}) for an arbitrary GF in terms of Chebyshev polynomials. Section \ref{2Dcase} proceeds with 2D systems. First we set the stage and present a straightforward generalization of the 1D calculation, which yields an analog of Eq.~(\ref{throughU}) for the 2D case: the Eq.~(\ref{2DGF}). After that we derive useful and simple expressions for the local density of states (DOS) at different positions in a lattice with open boundary. In Section \ref{PertTheory} we proceed with a system in a uniform magnetic field. Here our goal is to go beyond the seminal results of [\onlinecite{Hatsugai93PRB}] and [\onlinecite{Belov1991}]  by deriving an explicit expression for the GF, which remains valid for arbitrary magnetic fields. For not too strong magnetic fields we derive analytical expressions for the GFs and assess their quality by comparison with the exact results. The subsequent two sections are devoted to topological superconductors. In Section \ref{Kitaev_chain_model} we apply our method in order to derive Eq.~(\ref{GkmResult}), which is an explicit formula for an arbitrary GF of a Kitaev chain of finite length. This is one of the central results of the present work. In Section \ref{pwave} we turn to 2D $p$-wave superconductor on a lattice of finite size. Using the previously developed approach we find non-local corner states akin to edge zero modes of open Kitaev chains and discuss their possible applications. Finally, in Section \ref{3Dcase} we present some previously unknown results for the 3D tight-binding lattice. A Conclusions section offers a short summary of our findings.

\section{Basics: 1D tight-binding chain}
\label{Basics}

We begin our exposition with the simplest case -- the spinless 1D tight-binding chain of length $N$ with the Hamiltonian 
\begin{eqnarray}      \label{H0x}
 H_{\rm tb} = \sum_{i=1}^N \epsilon \, c_i^\dag c_i + \sum_{i=1}^{N-1} \gamma \, c_{i}^\dag c_{i+1} + \gamma^* \, c^\dag_{i+1} c_{i}  \, ,
\end{eqnarray} 
where $\epsilon$ is the uniform energy on each site and $\gamma$ is the (in general complex) hopping amplitude between the sites. Our primary goal is the Matsubara GF of the form 
\begin{eqnarray}       \label{Matsunm}
 g_{k m}(\tau) = \langle T_\tau \, c_k(\tau) \, c^\dag_m(0) \rangle \, ,
\end{eqnarray}
where $T_\tau$ is the imaginary time ordering operator. The most straightforward way to evaluate it is the direct diagonalization in terms of new operators $d_l$, 
\begin{eqnarray}                   \label{Sil}
 c_i = \sqrt{\frac{2}{N+1}} \sum_{l=1}^N \sin (q_l i) \, d_l = S_{i l}  \, d_l  \, ,
\end{eqnarray}
where the momenta underlie the following quantization condition:
\begin{eqnarray}
q_l = \frac{\pi l}{N+1} \, , \, \, \, \, \,  1 \le l \le N \, .
\end{eqnarray}
The diagonalized Hamiltonian is then 
\begin{eqnarray}      \label{}
 H = 2 \gamma \sum_{l=1}^N \cos(q_l) \, d^\dag_l \, d_l \, ,
\end{eqnarray} 
and the particle dispersion is obviously $E_l = 2 \gamma \cos(q_l)$. In this representation the evaluation of the GF is straightforward and leads to
\begin{eqnarray} 
 g_{k m}(i \omega_n) = \sum_{p=1}^N S_{k p} \, \frac{1}{i \omega_n - E_p} \, S_{p m} \, .
\end{eqnarray}
Thus the result is
\begin{eqnarray}    \label{unknown1}
  && 
  g_{k m}(i \omega_n) =   \frac{2}{N+1} 
  \nonumber \\     &\times&
  \sum_{p=1}^N \frac{ \sin [ \pi k p/(N+1)] \, \sin [\pi p m /(N+1)] }{i \omega_n - 2 \gamma \cos [\pi p/(N+1)]}  \, .
\end{eqnarray}
The remaining sum can be evaluated and one obtains
\begin{eqnarray}                \label{throughU}
  g_{k m}(i \omega_n) 
  = 
   \frac{ U_{k-1} \left( \frac{i \omega_n}{2\gamma} \right) \,  U_{N-m} \left( \frac{i \omega_n}{2\gamma} 
  \right)}{ \gamma U_{N} \left( \frac{i \omega_n}{2\gamma} \right)} \, ,
\end{eqnarray}
where $U_k(x)$ denote the Chebyshev polynomials of the second kind.\cite{gradshteyn2007, gogolin2013lectures,PeresStauberDosSantos09} 
This expression holds for $k \le m$, for $m > k$ we just have to interchange the indices. An alternative calculation can be performed using the functional integral formalism. The partition function is given by
\begin{eqnarray}
\label{funin1}
 Z = \int {\cal D}[c^\dag,c] e^{i S_1} \, , \, \, \, \,
  S_1 = \frac{1}{\beta} \sum_{i \omega_n} c^\dag(i \omega_n) \, {\cal A} \, c(i \omega_n) \, ,
\end{eqnarray}
where 
\begin{eqnarray}					\label{compfields}
c^\dag(i \omega_n) = (c_1^\dag(i \omega_n), \dots, c_N^\dag(i \omega_n))
\end{eqnarray}
are {\it composite fields} and the action kernel is given by an $N \times N$ matrix (from now on we concentrate on purely real $\gamma$, the case of generic tunnelling amplitude can be analyzed in exactly the same way)
\begin{eqnarray}					\label{Amatrix}
{\cal A}  = \left(
\begin{array}{cccccc}
i \omega_n & - \gamma & 0 & \ddots & 0 & 0 \\
- \gamma & i \omega_n & - \gamma & \ddots & 0 & 0 \\
0 & - \gamma & i \omega_n & \ddots &0 & 0 \\
\ddots & \ddots & \ddots & \ddots & \ddots & \ddots \\
0 & 0 & 0 & \ddots & i \omega_n &  - \gamma \\
0 & 0 & 0 & \ddots & - \gamma & i \omega_n 
\end{array} \right) \, .
\end{eqnarray}
The matrix (\ref{throughU}) of all possible GFs is just the inverse ${\cal A}^{-1}$. It can be calculated by the procedure proposed in [\onlinecite{USMANI199459}]. The resulting recurrence relations have solutions in terms of Chebyshev polynomials and one immediately obtains (\ref{throughU}). We note in passing that similar methods can be used to treat systems with periodic boundaries. For basic results in 1D see Appendix \ref{AppB}.

One practical application of the above result is the computation of the local density of states (DOS), which is found as the imaginary part of the local (at $k=m$) retarded GF. The latter is conveniently found from the analytically continued Matsubara GF via the substitution $i \omega_n \to \omega + i \delta$, where $\delta$ is a positive infinitesimal. Particularly interesting is the case of the edge site at $k=m=1$ or $k=m=N$ (from now on we use $\gamma$ as the energy unit), 
\begin{eqnarray}          			\label{DOSedge1D}
g_{\rm end}(i \omega_n) =  \lim_{N \to \infty} 
\frac{ U_{N-1} \left( \frac{i \omega_n}{2} 
  \right)}{U_{N} \left( \frac{i \omega_n}{2} \right)} 
 \nonumber \\
=   \frac{i \omega_n}{2} + \sqrt{\left( \frac{i \omega_n}{2} \right)^2 - 1} \, ,
\end{eqnarray}
where the evaluation of the limit can be done in accordance with the procedure outlined in [\onlinecite{Duran1999304}] (for the proper analytical continuation see Appendix \ref{AppeA}). We note in passing that in different applications this kind of GF is also referred to as {\it surface} or {\it boundary} GF. 

The same can be accomplished via the Dyson equation for the GF of the outmost chain site: 
\begin{eqnarray}               \label{Dyson1}
g^{-1}_{11}(N,i \omega_n) = g^{-1}_0(i \omega_n) - \gamma^2 \, g_{11}(N-1,i \omega_n) \, .
\end{eqnarray}
Here $g^{-1}_0(i \omega_n) = g^{-1}_{11}(1, i \omega_n) = i \omega_n - \epsilon$ is the reciprocal of the Matsubara GF of an individual uncoupled chain site. In the limit $N \to \infty$ we can set $g_{\rm end}(i \omega_n) = g_{11}(N,i \omega_n) = g_{11}(N-1,i \omega_n)$ and solve the corresponding equation.  The GF of the bulk site (at $k=m=N/2$) in the limit of the infinitely long chain 
$N \to \infty$ can also be computed by a version of the above Dyson equation. In this situation $2 g_{\rm end}(i \omega_n)$ plays the role of the self-energy and replaces $g_{11}(N-1,i \omega_n) $ in Eq.~(\ref{Dyson1}):\footnote{Obviously, the reversed procedure: from the bulk to the `surface' GF is as simple, see also [\onlinecite{VelevButler_2004}].}
\begin{eqnarray}					\nonumber
g^{-1}_{\rm bulk}(i \omega_n) = g^{-1}_0(i \omega_n) - 2 \gamma^2 \, g_{\rm end}(i \omega_n) \, .
\end{eqnarray}
This equation can be considered to be the simplest version of the {\it bulk-boundary correspondence}, often considered especially in the context of systems with topologically non-trivial band structures.\cite{EssinGurarie}

\section{2D tight binding system}
\label{2Dcase}
\subsection{Zero field case}

We construct a 2D system out of $M$ 1D systems of length $N$ arranged in parallel and coupled by the same matrix elements:
\begin{eqnarray}      \label{H1x}
 H_{\rm tb} 
= 
\sum_{n=1}^N \sum_{m=1}^M 
\epsilon \, c_{n, m}^\dag c_{n, m} + 
\gamma   \sum_{n=1}^{N-1} \sum_{m=1}^{M-1}
 c_{n, m}^\dag c_{n, m+1} 
\nonumber \\
+ 
 c^\dag_{n, m+1} c_{n, m}  
+
c_{n, m}^\dag c_{n+1, m} 
+ 
c^\dag_{n+1, m} c_{n, m} 
 \, .
\end{eqnarray} 
The partition function for this system can be written down in terms of a functional integral over $M$ composite fields, which are this time arrays of objects (\ref{compfields}):  
\begin{eqnarray}				\nonumber
Z = \int \left( \prod_{j=1}^M {\cal D} c_j^\dag \, {\cal D} c_j \right)\, e^{i S_2} \, , 
\end{eqnarray}
where
\begin{eqnarray}				\nonumber
&& S_2 =  \sum_{i \omega_n} \sum_{j=1}^M c_j^\dag(i \omega_n) \, {\cal A} \, c_j(i \omega_n) 
 \\ \nonumber
&+& 
\sum_{j=1}^{M-1} c_j^\dag(i \omega_n) \, \boldsymbol{\Gamma}^\dag c_{j+1}(i \omega_n)
  +  c_{j+1}^\dag(i \omega_n) \, \boldsymbol{\Gamma} c_j(i \omega_n) \, . 
\end{eqnarray}
Here ${\cal A}$ is as defined in (\ref{Amatrix}) and $\boldsymbol{\Gamma}=-\mbox{diag}(\gamma, \dots, \gamma)$  is the diagonal matrix of rank $N$ coupling the chains. By a repeated integration over the fields $c_1, \dots, c_{M-1}$ one obtains
\begin{eqnarray}				\nonumber
Z = \int {\cal D} c_M^\dag \, {\cal D} c_M\, e^{i S_2'} \, , 
\end{eqnarray} 
with the action
\begin{eqnarray}				\nonumber
S_2' =  \sum_{i \omega_n} c_M^\dag(i \omega_n) \, ( {\bf g}_1^{-1} -  \boldsymbol{\Gamma}^\dag \, {\bf g}_{M-1} \,
  \boldsymbol{\Gamma} ) \, c_M(i \omega_n) \, .
\end{eqnarray}
Here ${\bf g}_1 = {\cal A}^{-1}$ denotes the matrix of all GFs for an individual 1D chain with length $N$, given in Eq.~(\ref{throughU}). Thus the GF matrix for the sites at the edge of the system is found from the recurrence relation
\begin{eqnarray}				 \label{Dyson2}
{\bf g}_M^{-1} = {\bf g}_1^{-1} -  \boldsymbol{\Gamma}^\dag \, {\bf g}_{M-1} \,
  \boldsymbol{\Gamma}  \, ,
\end{eqnarray}
with the initial value ${\bf g}_0=0$.
This is a direct generalization of the relation (\ref{Dyson1}) and also has the form of a Dyson equation. 

As is shown in Ref.~[\onlinecite{MajoranaAK2016}] Eq.~(\ref{Dyson2}) allows for an explicit solution in terms of matrix polynomials of Chebyshev type.\cite{FaberTichy2010} Here we take another route and write the unknown GF as a quotient of two matrices ${\bf g}_M = P_M \, Q_M^{-1}$. Then the recurrence relation can be split into two: 
\begin{eqnarray}		\label{subs1}
P_{M+1} &=&  {\bf \Gamma}^{-1} \, Q_M \, , \, \, \, \, 
\\ \nonumber
Q_{M+1} &=& {\bf g}_1^{-1} \, {\bf \Gamma}^{-1} \, Q_M - {\bf \Gamma}^\dag \, P_M \, ,
\end{eqnarray}
which can immediately be solved in terms of powers of a matrix $R$:
\begin{eqnarray}                   \label{matrixSol1}
 \left( 
\begin{array}{c}
P_{M} \\
Q_{M}
\end{array}
\right) = R^{M} \,  \left( 
\begin{array}{c}
P_{0} \\
Q_{0}
\end{array}
\right) \, , 
\end{eqnarray}
where 
\begin{eqnarray}				\nonumber
R = \left(
\begin{array}{cc}
0 & {\bf \Gamma}^{-1} \\
- {\bf \Gamma}^\dag & {\bf g}_1^{-1} \, {\bf \Gamma}^{-1}
\end{array}
\right) \, .
\end{eqnarray}
Obviously the initial conditions are $P_0 = 0$ and $Q_1 = 1$. 
Let us now assume that $\gamma$ is purely real (this assumption considerably simplifies calculations and is not restrictive in any way) and our energy unit. Then 
\begin{eqnarray}  				\label{coolGamma}				\nonumber
\boldsymbol{\Gamma} =  \boldsymbol{\Gamma}^{-1} =  \boldsymbol{\Gamma}^\dag = 1
\end{eqnarray}
and 
\begin{eqnarray}			 \label{simpleR}
 R = \left(
\begin{array}{cc}
0 &1  \\
- 1 & {\cal A}
\end{array}
\right) \, .
\end{eqnarray}
We would like to find $T$ and a diagonal $R_0$, so that $R= T \, R_0 \, T^{-1}$. To that end we need the eigenvalues and -vectors of the matrix $R$. Its characteristic equation for the eigenvalues $\lambda$ reads
\begin{eqnarray} 				\label{eigenCheb}				\nonumber
  \mbox{det} \left[
 (1 + \lambda^2) \, 1
  - \lambda {\cal A}   
\right] 
= \lambda^N \, U_N\left( \frac{1 + \lambda^2}{2 \lambda} - \frac{i \omega_n}{2} \right) = 0 \, .
\end{eqnarray}
This equation can be solved using the trigonometric representation of the Chebyshev polynomials. The $2 N$ different solutions are given by 
\begin{eqnarray}				\nonumber
 \label{eigenvalues}
 \lambda_{k 1,2} = \left[ \frac{i \omega_n}{2} - \cos \left( \frac{\pi k}{N+1} \right) \right] 
 \nonumber \\
 \pm
 \sqrt{\left[ \frac{i \omega_n}{2} - \cos \left( \frac{\pi k}{N+1} \right) \right]^2 -1} \, , 
\end{eqnarray}
where $1 \le k \le N$. Therefore $R_0=$diag$( \lambda_{j 1}, \lambda_{j 2})$.   
The matrix $T$ can be written down in terms of eigenvectors $v_{1,\dots, N}$ of the action matrix ${\cal A}$ for a 1D chain:  
\begin{eqnarray}				\nonumber
T = \left(
\begin{array}{cc}
 A & B \\
 C & D
\end{array}
\right) 
\end{eqnarray}
with 
\begin{eqnarray}				\nonumber
 C &=& D = (v_1, \dots, v_N) \, , \, \, A=  (v_1/\lambda_{1 1}, \dots, v_N/\lambda_{N 1}) \, ,
 \nonumber \\
 B &=& (v_1/\lambda_{1 2}, \dots, v_N/\lambda_{N 2}) \, .
\end{eqnarray}
$v_j$ constitute a self-inverse symmetric matrix $V$, which is essentially a square root of a unity matrix, 
\begin{eqnarray}					\label{V}				\nonumber
V =  (v_i)_j = \sqrt{\frac{2}{N+1}} \sin \left( \frac{\pi \, i \, j}{N+1} \right) \, .
\end{eqnarray}
In fact, this result is already contained in the diagonalization transformation (\ref{Sil}). Computation of powers of $R$ is now straightforward and is just 
\begin{eqnarray}				\label{powersofR}
   &&	
   R^{q} = T \,  \mbox{diag}( \lambda^{q}_{j 1}, \lambda^{q}_{j 2}) \, T^{-1} 
\nonumber \\
 &=& 
= \left(
\begin{matrix}
- v_{i j} \, U_{q-2} (\epsilon_j)  \, v_{j k} & 
v_{i j} \, U_{q-1}(\epsilon_j)  \, v_{j k}
\\
- v_{i j} \, U_{q-1}(\epsilon_j)  \, v_{j k}  & 
v_{i j} \, U_{q} (\epsilon_j)  \, v_{j k}
\end{matrix}
\right) \, , 
\end{eqnarray}
where 
\begin{eqnarray}                \label{epsilonArgument}
  \epsilon_j = \frac{ i \omega_n}{2} - \cos \left( \frac{ \pi j}{N+1} \right) \, .
\end{eqnarray}
Due to the special property of the recurrence initial conditions only the right column of the result (\ref{powersofR}) is important for the GF computation. 
Plugging this back into (\ref{matrixSol1}) and computing the ratio of $P_{M} \, Q_M^{-1}$ we obtain the final result: 
\begin{eqnarray}			 \label{unknown2}
{\bf g}(i \omega_n) = V \, {\cal B} \, V \, , \, \, \, 
{\cal B} = \mbox{diag} \left(
 \frac{U_{M-1} (\epsilon_j)}{U_{M} (\epsilon_j)}
\right) \, .
\end{eqnarray}
In Eq.~(\ref{unknown2}) one can immediately recognize the 1D result  (\ref{unknown1}). On the other hand, from (\ref{unknown1}) and (\ref{throughU}) follows the identity
\begin{eqnarray}				\nonumber
\frac{U_{M-1}(\epsilon)}{U_{M}(\epsilon)} &=& 
\frac{2}{M+1} \sum_{p=1}^M \frac{\sin \left( \frac{\pi p}{M+1} \right) \sin \left( \frac{\pi p}{M+1} \right)}{2 \epsilon - 2 \cos \left( \frac{\pi p}{M+1} \right)} 
\nonumber \\
&=&  \frac{1}{2} \sum_{p=1}^M \frac{w_{1p} \, w_{p1}}{ \epsilon - \cos \left( \frac{\pi p}{M+1} \right)} \,, 
\end{eqnarray}
where 
\begin{eqnarray}					\label{W}				\nonumber
W =  (w_i)_j = \sqrt{\frac{2}{M+1}} \sin \left( \frac{\pi \, i \, j}{M+1} \right) \, .
\end{eqnarray}
Therefore the GF between the sites $(i,p)$ and $(j,q)$ is 
\begin{eqnarray}  \nonumber
g_{(ip),(jq)}(i \omega_n) &=&
 \sum_{r=1}^M  \sum_{k=1}^N
 \frac{ v_{i k} v_{k j} \, w_{p r} \, w_{r q}}{\frac{i \omega_n}{2} - \cos \left( \frac{\pi k}{N+1} \right)
- \cos \left( \frac{\pi r}{M+1} \right)}
\nonumber \\      					\label{2DGF}
&=&
\sum_{k=1}^N v_{i k} v_{k j}
 \frac{ U_{p-1} \left( \epsilon_k \right) \,  U_{M-q} \left( \epsilon_k 
  \right)}{  U_{M} \left( \epsilon_k \right)} \, .
\end{eqnarray}

\begin{widetext}
As an application we compute the DOS at different points in the lattice, see Fig.~\ref{Fig1}: (a) at the corner site of the lattice; (b) at the site on the edge of the system far away from the corners, which we call {\it edge bulk} (eb) site; (c) at the bulk far away from the edges. While an analytical result for (c) exists and is reported in [\onlinecite{2D_DOS_old}], the situations (a) and (b) have not yet been considered.

\begin{figure}[h]
  \centering    \includegraphics[width=0.7\columnwidth]{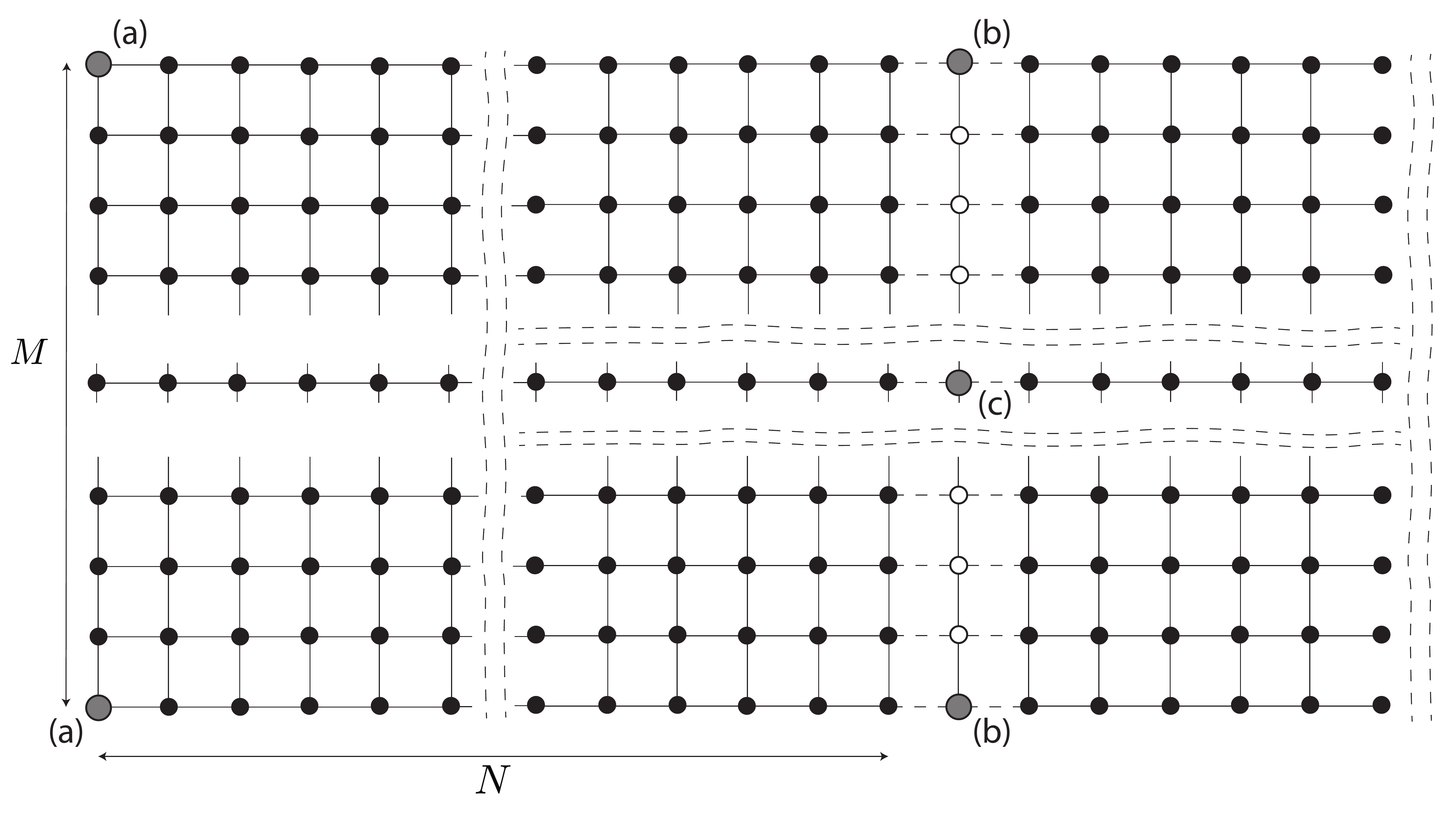}
  \caption{  \label{Fig1}
  2D tight binding lattice. (a) represent the corner site, (b) the edge bulk site and (c) is the true bulk site.
}
\end{figure}
\end{widetext}

First we concentrate on the corner site $(1,1)$ of our lattice for $M \to \infty$ and $N \to \infty$. Taking the latter limit amounts to a replacement of the quotient of the polynomials in (\ref{unknown2}) by (\ref{DOSedge1D}) with appropriate arguments. Using the result of the analytical continuation given in Appendix \ref{AppeA} and replacing the sum in  (\ref{unknown2}) by an integral we then obtain the following result for positive energies $\omega > 0$:
\begin{eqnarray}                                    \label{corner}
\nu_{\rm corner}(\omega) = \frac{2}{\pi} \int\limits_{\omega/2 - 1}^1 d y \sqrt{(1 - y^2)[1 - (\omega/2-y)^2]}  
\, .
\end{eqnarray}
With the help of a similar procedure we can compute the DOS at a site $(1,N/2)$ in the middle of the system edge, at the edge bulk site. Here in the limit $N \to \infty$ we obtain  
\begin{eqnarray}					\label{eb1}
\nu_{\rm eb}(\omega) = \frac{1}{\pi} \int_{\omega/2 - 1}^1 d y \sqrt{\frac{1 - (\omega/2-y)^2}{1 - y^2}}  
\, .
\end{eqnarray}

Finally, the genuine bulk GF can be computed using the version of the bulk-boundary correspondence condition following from Eq.~(\ref{Dyson2}):
\begin{eqnarray}				 \label{BB2D}				\nonumber
{\bf g}_{\rm bulk}^{-1} = {\bf g}_1^{-1} - 2\boldsymbol{\Gamma}^\dag \, {\bf g} \,
  \boldsymbol{\Gamma}  \, .
\end{eqnarray}
This equation describes a 1D system of length $N$ with the GF ${\bf g}_1$ (the chain with open circles on Fig.~\ref{Fig1}), which is coupled to two identical systems of sizes $N \times M$ with edge GFs ${\bf g}$.\footnote{These systems can, of course, in general be different.} Since 
\begin{eqnarray}				\nonumber
{\bf g}_1^{-1} = {\cal A} = V \, C \, V \, ,
\end{eqnarray}
where $C= \mbox{diag} (i \omega_n - 2 \cos[\pi k/(N+1)])$ we thus obtain 
\begin{eqnarray} \nonumber 
{\bf g}_{\rm bulk} = V \mbox{diag} \left[ 
 i \omega_n - 2 \cos\left(\frac{\pi k}{N+1}\right) -2  \frac{U_{M-1} (\epsilon_k)}{U_{M} (\epsilon_k)}
\right]^{-1}  V \, .
\end{eqnarray}
Using this result we derive an alternative expression for the edge bulk DOS of an infinitely large system:
\begin{eqnarray}				\label{eb2}
\nu_{\rm eb}(\omega) = \frac{1}{\pi} \int_{\omega/2 - 1}^1 d y \sqrt{\frac{1 - y^2}{1 - (\omega/2-y)^2}}  
\, ,
\end{eqnarray}
which yields exactly the same result as Eq.~(\ref{eb1}). For the genuine bulk DOS we then obtain the known result \cite{2D_DOS_old}
\begin{eqnarray}                                \label{bulkDOS}
\nu_{\rm bulk}(\omega) = \frac{1}{2 \pi} \int\limits_{\omega/2 - 1}^1 \frac{d y }{\sqrt{(1 - y^2)[1 - (\omega/2-y)^2]} } 
\, .
\end{eqnarray}
The expressions (\ref{corner}), (\ref{eb1}), (\ref{eb2}), (\ref{bulkDOS}) are valid for positive energies $0<\omega <2$, for $\omega>2$ (outside of the band) the DOS is zero in all three cases. We would like to remark that all four integrals can be expressed in terms of elliptic functions. We abstain from doing so as it does not produce any added value.

\subsection{2D lattice in magnetic field}				\label{PertTheory}

\begin{figure}
  \centering    \includegraphics[width=0.9\columnwidth]{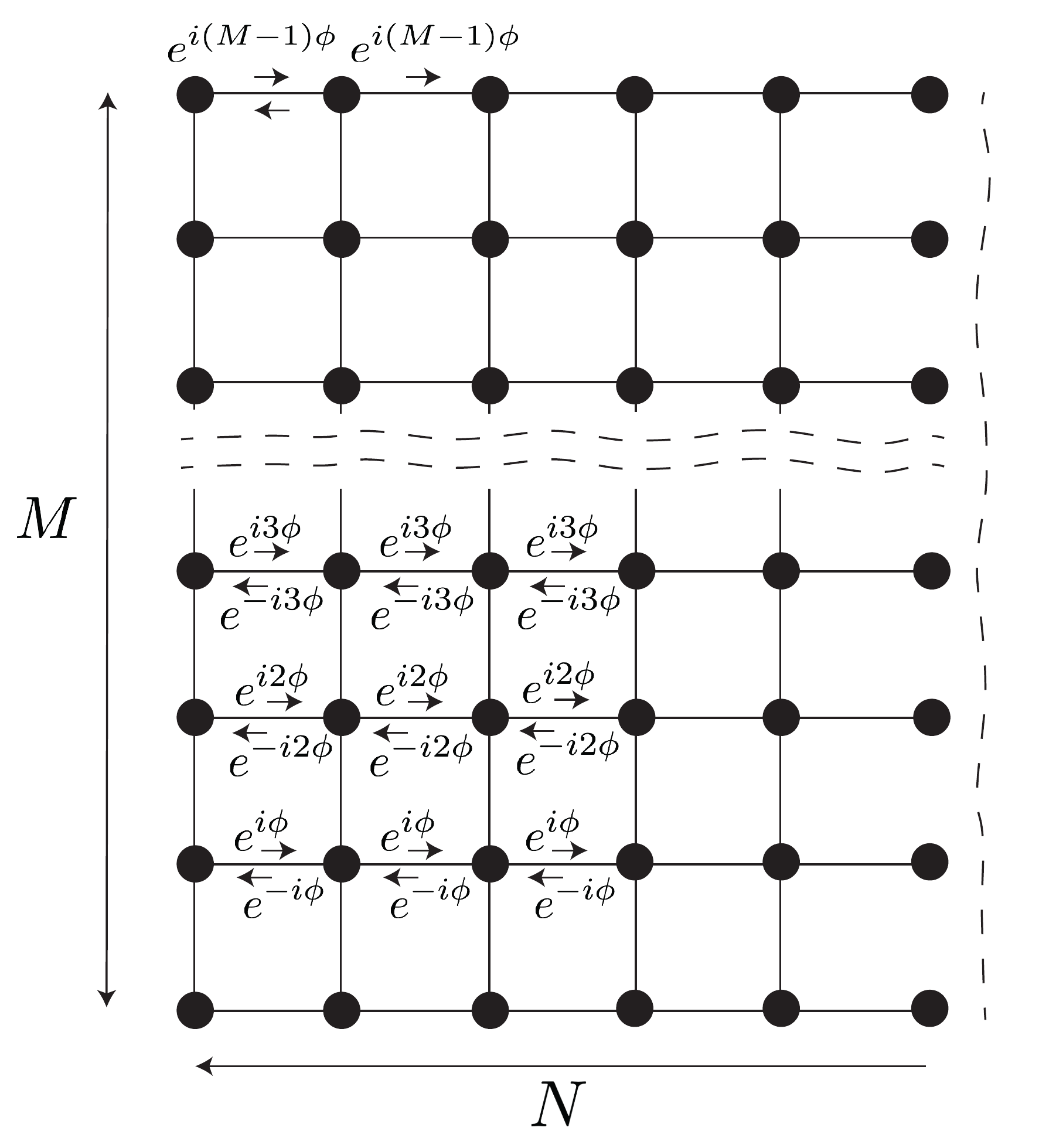}
  \caption{  \label{Fig2}
  2D tight binding lattice in magnetic field pointing perpendicular to the lattice plane. Every amplitude $\gamma$ describing a tunnelling process from left to right is to be supplemented by a factor $e^{i m \phi}$ and that of the opposite direction by a factor $e^{-i m \phi}$.}
\end{figure}

In order to include magnetic field into the model (\ref{H1x}) we use Landau gauge. 
We apply the Peierls substitution in the form depicted in Fig.~\ref{Fig2}.\cite{Peierls33}
In this kind of geometry 1D systems of length $M$ are coupled by `bare' tunnelling amplitudes $\gamma$, which do not contain magnetic field dependent phases. On the other hand, the tunnelling amplitude within each of such 1D chains is dressed by factors $e^{ \pm i m \phi}$, 
where $0<m<M-1$ is the chain index. The phase is defined as $\phi = B a_0^2/\Phi_0$, where $B$ is the field magnitude, $a_0$ is the lattice constant and $\Phi_0 = h/2e$ is the magnetic flux quantum. 

The GF for particles in each of the $M$ chains is in analogy to (\ref{Amatrix}) given by the inverse of the corresponding $N \times N$ action matrix:
\begin{eqnarray}					\label{Ammatrix}				\nonumber
{\cal A}_m  = \left(
\begin{array}{cccccc}
i \omega_n & - \gamma  e^{i m \phi}  & 0 & \ddots   \\
- \gamma  e^{-i m \phi}  & i \omega_n & - \gamma  e^{i m \phi} & \ddots  \\
0 & - \gamma e^{-i m \phi}  & i \omega_n & \ddots   \\
\ddots & \ddots & \ddots & \ddots & \\
\end{array} \right) \, .
\end{eqnarray}
Unsurprisingly, the eigenvalues $\epsilon_k = i \omega_n/2 - \cos[ \pi k/(M+1)]$ ($1<k<M$) of this matrix do not depend on $\phi$ and the matrix of its eigenvectors is given by a still self-inverse matrix
\begin{eqnarray}				 \label{Eigenvectors}				\nonumber
 U_{k l} = \sqrt{\frac{2}{N+1}} \sin \left( \frac{\pi k l}{N+1} \right) \, e^{- i  (k-l) m \phi} \, ,
\end{eqnarray}
so that ${\cal A}_m = U \, B \, U$, where $B=\mbox{diag}(2 \epsilon_n)$. 
For the GF, which is a direct generalization of  (\ref{throughU}) one then obtains
\begin{eqnarray}                \label{throughUB}				\nonumber
  && g_{k l}(i \omega_n) = [{\bf g}_1(i \omega_n, m)]_{k l} = 
 e^{- i m \phi (k - l)} 
 \\  \nonumber
  &\times&
   \frac{1}{ \gamma U_{N} \left( \frac{i \omega_n}{2\gamma} \right)}
  \left\{ \begin{array}{cc}
 U_{k-1} \left( \frac{i \omega_n}{2\gamma} \right) \,  U_{N-l} \left( \frac{i \omega_n}{2\gamma} 
  \right) \, ,
& k \le l \\
 U_{l-1} \left( \frac{i \omega_n}{2\gamma} \right) \,  U_{N-k} \left( \frac{i \omega_n}{2\gamma} 
  \right)  \, ,
  & k > l 
  \end{array} \right. \, .
\end{eqnarray}
Such 1D systems are coupled to each other by the same matrices $\boldsymbol{\Gamma}$ as in the field free case. Therefore the recurrence relation for the GFs of the edge chain reads
\begin{eqnarray}				\nonumber
 \label{finiteFi}
 {\bf g}_{M}^{-1}(i \omega_n) = {\bf g}_1^{-1}(i \omega_n, (M-1) \phi) - \boldsymbol{\Gamma}^\dag {\bf g}_{M-1} (i \omega_n) \boldsymbol{\Gamma} \, .
\end{eqnarray}
Its solution can be constructed in terms of a matrix $R_s$ defined as
\begin{eqnarray}				\nonumber
R_s = \left(
\begin{array}{cc}
0 & 1 \\
- 1 & {\cal A}_m
\end{array}
\right) =  \left(
\begin{array}{cc}
0 & 1 \\
- 1 & {\bf g}_1^{-1} [i \omega_n, (s-1) \phi]
\end{array}
\right) \, .
\end{eqnarray}
In this notation the above recurrence relation reads
\begin{eqnarray}				\nonumber
\left( 
\begin{array}{c}
P_{M} \\
Q_{M}
\end{array}
\right) = R_{M} \,  \left( 
\begin{array}{c}
P_{M-1} \\
Q_{M-1}
\end{array}
\right) \, ,
\end{eqnarray}
where as in the previous subsection $ {\bf g}_{M}(i \omega_n) = P_M Q_M^{-1}$,  $P_0=0$ and $Q_0=1$. Its solution obviously is 
\begin{eqnarray}				\nonumber
\left( 
\begin{array}{c}
P_{M} \\
Q_{M}
\end{array}
\right) =  \left( \prod_{s=M}^1 R_s \right) \left( 
\begin{array}{c}
P_{0} \\
Q_{0}
\end{array}
\right) \, .
\end{eqnarray}
We use the following substitution: 
\begin{eqnarray}				\nonumber
\Delta_s
=
R_0^{-1}R_s
=
1+\epsilon_s \, ,
 \, \, \, 
 \epsilon_s
=
\left(
\begin{matrix}
0 & D_s\\
0 &0 \\
\end{matrix}
\right) \, ,
\end{eqnarray}
where $R_0$ is the matrix  (\ref{simpleR}) of the field-free case and
\begin{eqnarray}			\nonumber 
D_{s+1}
=
\left(
\begin{matrix}
0 & e^{is\phi}-1 & 0 & 0 &\ddots\\
e^{-is\phi}-1 & 0 & e^{is\phi}-1  & 0 & \ddots\\
0 & e^{-is\phi}-1 & 0 & e^{is\phi}-1  & \ddots\\
0 & 0 & e^{-is\phi}-1 & 0 & \ddots \\
\ddots & \ddots & \ddots & \ddots & \ddots  \\
\end{matrix}
\right)
\, .
\end{eqnarray}
With this notation the above matrix product reads
\begin{eqnarray} 				\nonumber
 && R_M \, R_{M-1} \, \cdots \, R_2 \, R_1
 = R_0 \Delta_M R_0 \Delta_{M-1} \cdots R_0 \Delta_1
\nonumber \\
&=&
R_0 (1 + \epsilon_M) R_0 (1 + \epsilon_{M-1}) \cdots R_0 (1 + \epsilon_1) \, .
\end{eqnarray}
This representation is very useful for expansion in small fields as $D_{s+1}$ and thus $\epsilon_{s+1}$ are objects of the order $s \phi$. Therefore the expansion to the order $(M \phi)^2$ is given by
\begin{eqnarray}				\nonumber
&& R_M \, R_{M-1} \, \cdots \, R_2 \, R_1 =
R_0^{M}
+
\sum_{s=2}^M
R_0^{M-s+1} \epsilon_s R_0^{s-1}
\nonumber \\
&+&
\sum_{i,j=2,\, \,  j >i}^M
R_0^{M-j+1} \epsilon_j R_0^{j-i} \epsilon_i R_0^{i-1}
+
\mathcal{O}(\epsilon^3)
\,.
\end{eqnarray}
In the next step we keep only terms linear in $\epsilon_s$. Since
\begin{eqnarray}				\nonumber
 R_0^{M-s+1} \epsilon_s R_0^{s-1}
= 
\left(
\begin{array}{cc}
\dots &
- V \, {\bf U}_{M-s-1} \, V \,  D_s \, V \,  {\bf U}_{s-1} \, V
\\
\dots
&
- V \, {\bf U}_{M-s} \, V \, D_s \, V \, {\bf U}_{s-1} \, V
\end{array}
\right) \, ,
\end{eqnarray}
where ${\bf U}_k = \mbox{diag} \, U_k(\epsilon_j)$ denotes a diagonal matrix containing Chebyshev polynomials of the argument $\epsilon_j$, $1<j<N$ as defined in Eq.~(\ref{epsilonArgument}), using the identity $U_{M-1} U_{M-s} - U_{M-s-1} U_M = U_{s-1}$ we obtain the following result for the GF on the edge of our system:
\begin{eqnarray}							\label{Pert1st}
 && {\bf g}(i \omega_n, \phi) = {\bf g}_0(i \omega_n,0) + {\bf g}_1(i \omega_n,\phi) + \dots 
 \\ \nonumber
&=&  {\bf g}(i \omega_n,0)  + \sum_{s=2}^M
V \, {\bf U}_{s-1} \,  {\bf U}^{-1}_M \, V \, D_s \, V \,  {\bf U}_{s-1} \,  {\bf U}^{-1}_{M} \, V 
+ \dots  
\end{eqnarray}
The form of this correction allows for an interesting and useful interpretation. In accordance with Eqs.~(\ref{throughU}) and (\ref{2DGF}) one can consider the factors $V \, {\bf U}_{s-1} \,  {\bf U}^{-1}_M \, V$ as being the GFs for the lattice nodes located in the rows $m=s$ and $m=M$, see Fig.~\ref{Fig2}. So the first factor from the right describes particle propagation from the row $m=M$ to the row $m=s$ where the particle `feels' the magnetic field, picks up the factor $D_s$ and after that propagates back to the row with index $M$ thereby correcting the field free result ${\bf g}(i \omega_n,0)$. According to this scheme the second order contribution can be found to be given by the following expression:
\begin{eqnarray}
&& {\bf g}_2(i \omega_n,\phi) =  \sum_{s, s' =2, s>s'}^M
 \left( V \,  {\bf U}_{s-1}  {\bf U}^{-1}_M \, V \right)
 \nonumber \\				\nonumber
 &\times&
 D_s  \left( V \, \left[ {\bf U}_{s-1} {\bf U}_{M-s'} {\bf U}^{-1}_M - {\bf U}_{s-s'-1} \right] V \right)
 \nonumber \\				\nonumber
 &\times&
 \, D_s'  \left( V \,  {\bf U}_{s'-1}  {\bf U}^{-1}_M \, V \right) \, .
\end{eqnarray}
Here the particle travels to the row $s'$, picks up the phase induced by $D_{s'}$, travels to the row $s>s'$, picks up the second phase due to $D_s$ and after that returns back. In the similar way one can construct corrections of arbitrary order.

\subsection{Edge states: the local density of states}

\begin{figure}[h]
  \centering    \includegraphics[width=1.\columnwidth]{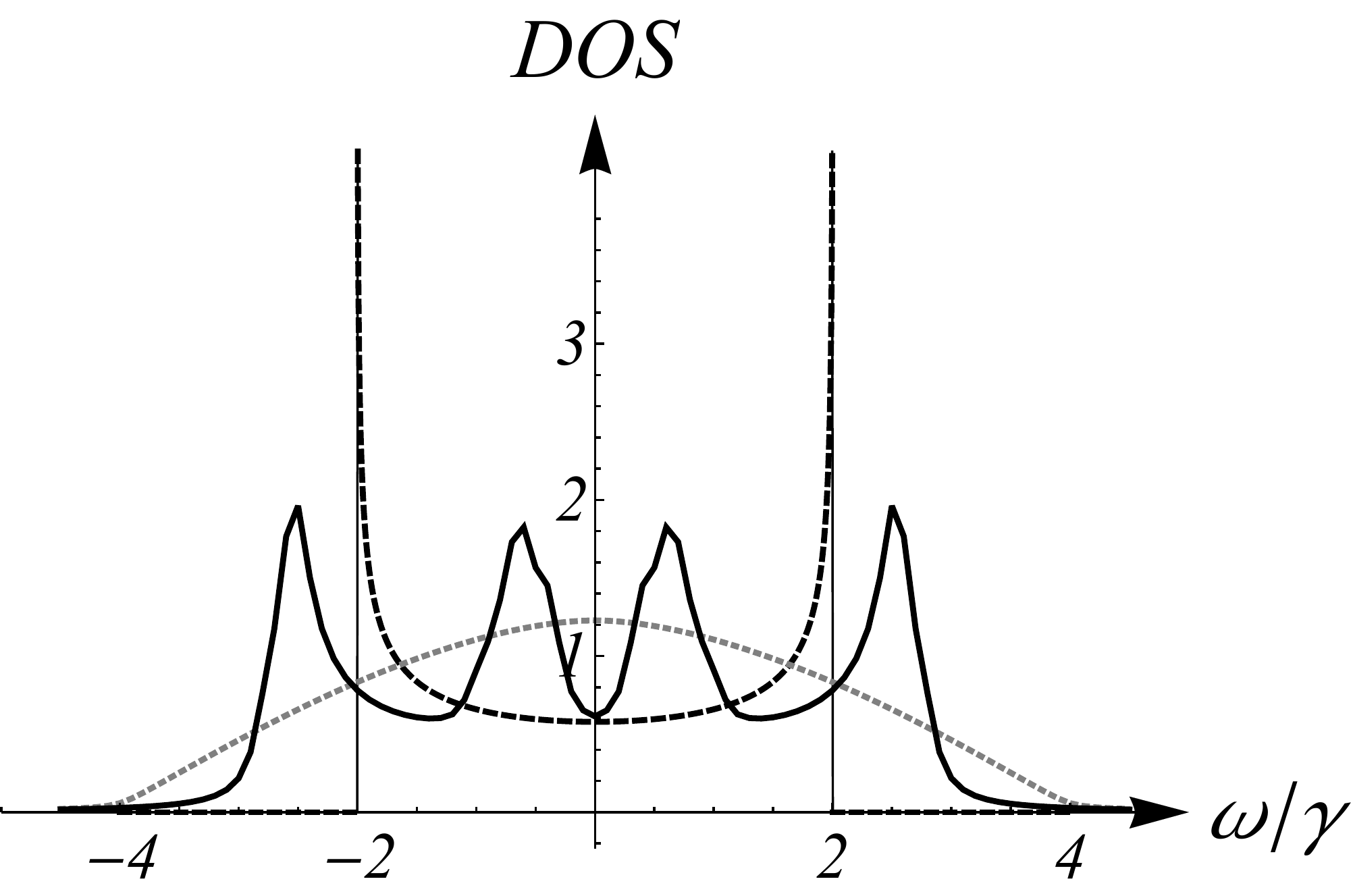}
  \caption{  \label{Fig2Dvs1D}
  Local density of states measured in arbitrary units at the edge of a non-interacting 2D tight-binding lattice far away from the lattice corners ({\it edge bulk} case) without the magnetic field (dotted line) and in the field with the strength $\phi/\phi_0 = 1.5$ (solid line). Both systems have dimensions $M=N=60$. Dashed line represents the local DOS in the bulk of an 1D system computed with the help of formula (\ref{1DlocalDosBulk}). The peak structure in the magnetic case indicated the presence of four Landau levels. 
  }
\end{figure}

It is known that in 2D systems subject to strong magnetic fields there are gapless edge states. It is interesting to recover them using the just developed technique. The simplest quantity is the local density of states (DOS), the spacial dependence of which is exemplarily plotted in Fig.~\ref{Fig2Dvs1D}.
A direct comparison of the edge DOS in the 2D case in presence of magnetic field with the one of a 1D system indicates the presence of edge channels with approximately 1D geometry.

\subsection{Edge states: the edge currents}
\label{edge-currents}

As we have seen in the previous subsection the local DOS is not sensitive to chirality of the edge states. In order to access this information a direct computation of particle currents is more appropriate. One can for example consider the current flowing between the sites $(n,M)$ and $(n+1,M)$ at the right outmost edge of a sample with dimensions $N$ and $M$. The corresponding operator is given by
\begin{eqnarray}                                   \label{current_def}
 J_n = - i \gamma \left( 
  c^\dag_{n,M} c_{n+1,M} - c^\dag_{n+1,M} c_{n,M} 
 \right) \, .
\end{eqnarray}
This quantity can be computed in the following way. Let us consider two 2D systems in the same magnetic field. The operators of one of them are $c_{n,m_1}$ and the operators of the other one $d_{n,m_2}$, where $1\le n \le N$ and $1 \le m_{1,2} \le M_{1,2}$. We gauge the field in such a way that the (for definiteness lower) edge $m_1=0$ of the first system does not carry phase factors and they grow in positive direction for growing index $m_1$ up to the value $(M_1 - 1) \phi$ in the opposite edge of the system. On the other hand, the (upper) edge of the second system carries the phase factor $-\phi$ and it grows into negative values up to $-M_2 \phi$ on the opposite edge of this subsystem. 

Let ${\bf g}$ be the GF of the sites on the lower edge of the system $1$ and ${\bf h}$ be the GF of the upper edge of the second system. Now we couple the systems by tunnelling. Then the effective action has the form
\begin{eqnarray}				\nonumber
 S = \frac{1}{\beta} \sum_{i \omega_n} ({\bf c}^\dag, {\bf d}^\dag) \left(
\begin{array}{cc}
{\bf g}^{-1} & {\Gamma} \\
{\Gamma} & {\bf h}^{-1}
\end{array}
\right)
\left(
\begin{array}{c}
{\bf c} \\ {\bf d}
\end{array}
\right) \, ,
\end{eqnarray}
where the notation $({\bf c}^\dag, {\bf d}^\dag)$ stands for the composite field $(c_1^\dag, c_2^\dag \dots c_N^\dag, d_1^\dag, d_2^\dag \dots d_N^\dag)$ and $\Gamma$ is a unit matrix times $- \gamma$.  We are interested in the GF between the neighboring lattice sites and we choose them to be of the kind $\langle T_\tau \, c_n(\tau) \, d^\dag_n(0) \rangle$ and 
$\langle T_\tau \, d_n(\tau) \, c^\dag_n(0) \rangle$. The difference between the two is precisely the current according to the definition (\ref{current_def}). These expectation values can be found via matrix inversion of the above action. 
Taking the difference of the off-diagonal components of the inverse matrix we obtain an array of currents
\begin{eqnarray}				\nonumber
 {\bf J} &=& - i \frac{\gamma}{\beta} \sum_{i\omega_n} \left[
  \langle {\bf c} {\bf d}^\dag \rangle 
  -  \langle {\bf d} {\bf c}^\dag \rangle 
  \right] 
 \\  \nonumber 
  &=& 
  - i \frac{\gamma^2}{\beta} \sum_{i\omega_n}  \left[
 {\bf h} (1 - \gamma^2 {\bf g}  {\bf h})^{-1} \, {\bf g} -  {\bf g} (1 - \gamma^2 {\bf h}  {\bf g})^{-1} \, {\bf h}   \right] \, .
\end{eqnarray}
This expression is odd with respect to exchange ${\bf g} \leftrightarrow {\bf h}$. This automatically yields currents of opposite signs through the links located at the same distance from the middle symmetry axis of the sample. The opposite signs for the currents on the opposite sample edges follow. Moreover, in the absence of the magnetic field the matrices ${\bf g}$ and ${\bf h}$ commute and the net current through the links vanishes. This is due to the fact that both subsystems are diagonalized by the same transformation.

\begin{figure}
  \centering    \includegraphics[width=0.85\columnwidth]{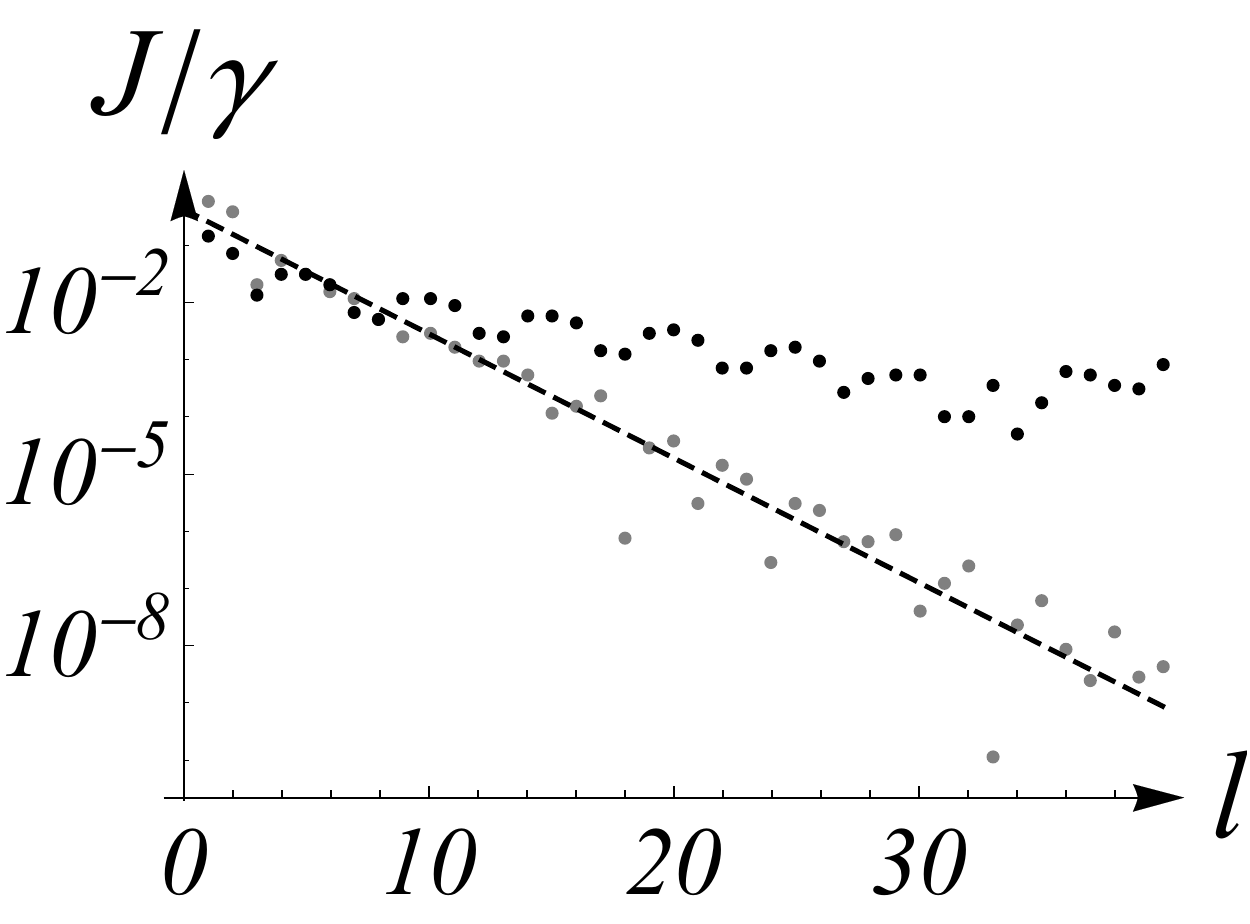}
  \caption{  \label{FigCurrentDecay}
  Logarithmic plot of the absolute values of the currents on the sites $l$ in the middle of an $80\times80$ lattice in a strong magnetic field $\phi/\phi_0=0.80 \pi$ at temperature $T/\gamma=0.1$ as a function of the distance from the edge. Grey dots represent the exact solution for the full current while the black dots show the results gained from the approximation derived in Section \ref{edge-currents}. The dotted line corresponds to the curve $f(l)=0.25 e^{-0.5(l-1)}$.}
\end{figure}

Numerical evaluation of the above expression is not difficult and one can conveniently discuss all features of the currents in the sample. Among other things one immediately verifies that the  currents decay exponentially with the distance from the sample edge, see Fig.~\ref{FigCurrentDecay}. We also find that there is no noticeable net field-driven depletion of charge even in the case of very strong fields. 

It turns out that the edge currents computation can be very conveniently performed using the perturbative expansion presented in Section \ref{PertTheory}. To the lowest order we can write ${\bf g} \approx {\bf g} _0+{\bf g} _1$ and ${\bf h} \approx {\bf h} _0+{\bf h} _1$ and use the result (\ref{Pert1st}). Defining ${\bf f}:= \mathbb{1}-\gamma^2{\bf g}_0 {\bf h}_0$ and taking advantage of the identity 
$\left[{\bf g} _0, {\bf h} _0 \right]=0$ one obtains
\begin{widetext}
\begin{eqnarray}				\nonumber
\frac{J_{i\omega_n}}{-i\gamma }
\approx
\left[{\bf h}_0{\bf f}^{-1},{\bf g}_1\right]
+
\gamma^2
{\bf h}_0{\bf f}^{-1}\left[{\bf g}_1,{\bf g}_0\right]{\bf h}_0{\bf f}^{-1} 
-({\bf g}_0 \leftrightarrow {\bf h}_0 \, , \, {\bf g}_1 \leftrightarrow {\bf h}_1)
\\ \nonumber
=
V
\left(
\left[H_0F^{-1},G_1\right]
+
\gamma^2
H_0F^{-1}\left[G_1,G_0\right]H_0F^{-1} 
\right)
V
-(G_0 \leftrightarrow H_0 \, , \, G_1 \leftrightarrow H_1)
\, ,
\end{eqnarray}
where we defined $J_{i\omega_n}$ as the energy-resolved current with the property $J=\frac{1}{\beta}\sum_{i\omega_n}J_{i\omega_n}$, as well as $G_{0,1}=V{\bf g}_{0,1}V$, $H_{0,1}=V{\bf h}_{0,1}V$ and $F=V{\bf f}V$.

Assuming  $\gamma$ to be the energy unit as before one arrives at an explicit expression for the current to first order in $e^{iM_1\phi}-1$ or $e^{iM_2\phi}-1$, whichever is larger:
\begin{eqnarray}
\frac{J_{i\omega_n}}{-i\gamma } &=&
\left(VJ_{i\omega_n}V\right)_{kl}
\approx
\delta_{k,l+1 \mod  2}
\frac{-4}{N+1}
\frac{
\sin\left(\frac{\pi}{N+1}k\right)
\sin\left(\frac{\pi}{N+1}l\right)
}{
\cos\left(\frac{\pi}{N+1}k\right)-\cos\left(\frac{\pi}{N+1}l\right)
}
\nonumber
\\				\nonumber
&\times&
\frac{
U_{M_2}(\epsilon_l) U_{M_2-1}(\epsilon_k)
-
U_{M_2}(\epsilon_k) U_{M_2-1}(\epsilon_l)
}{
[U_{M_2}(\epsilon_k) U_{M_1}(\epsilon_k) - U_{M_2-1}(\epsilon_k) U_{M_1-1}(\epsilon_k)]
\cdot
(k \leftrightarrow l)
}
\\  \nonumber
&\times&
\sum_{s=1}^{M_1}
U_{s-1}(\epsilon_k)
\sin[(s-1)\phi]
U_{s-1}(\epsilon_l)
- 
[M_1\leftrightarrow M_2\, ,\,  (s-1)\phi\rightarrow -s\phi] \, ,
\end{eqnarray}
\end{widetext}
where $\epsilon_k$ were defined in (\ref{epsilonArgument}) and the very last term denotes the term identical to the first one up to the indicated substitutions.
In the last computation step we have performed an index shift of $s$, which does not affect the index of the Chebyshev polynomials.

\begin{figure}
  \centering    \includegraphics[width=0.85\columnwidth]{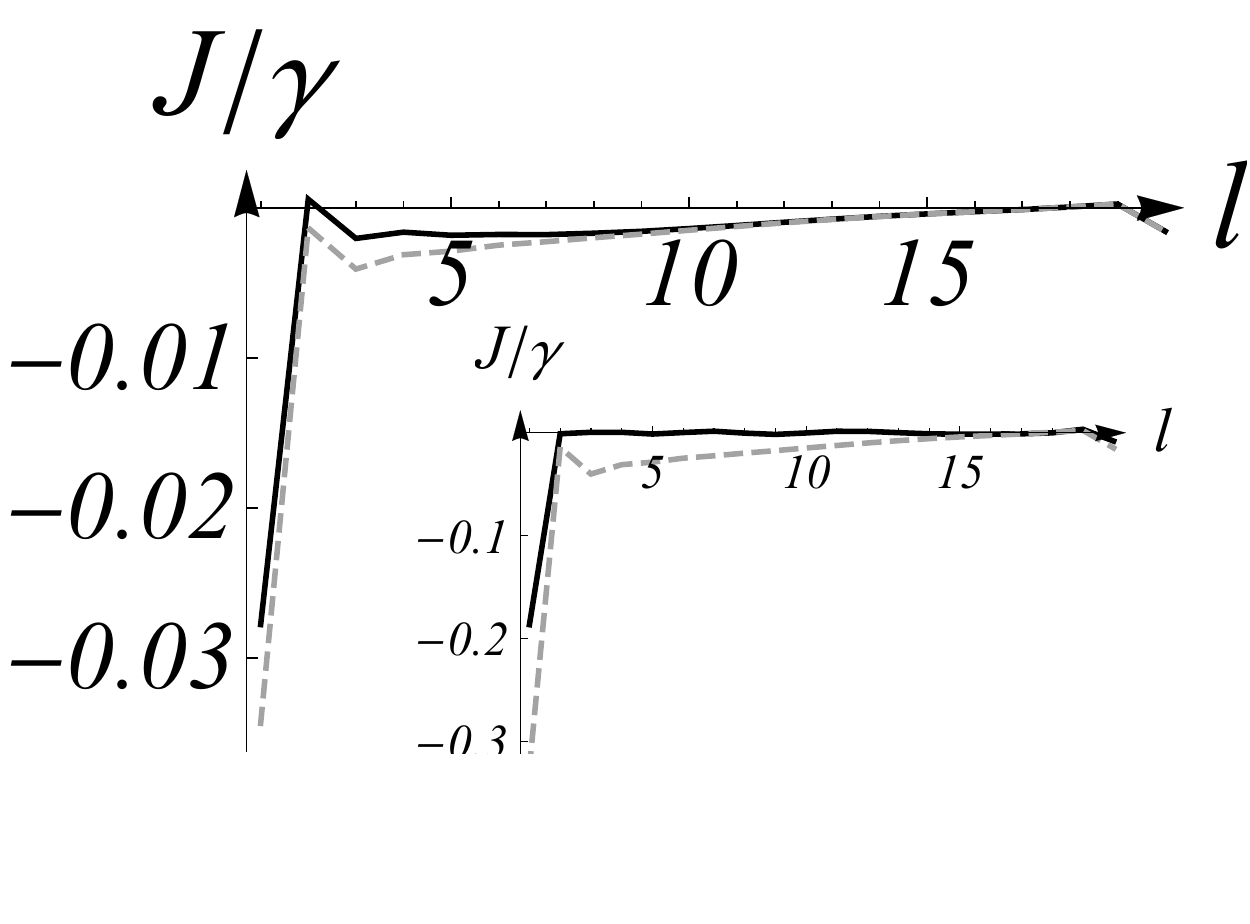}
  \caption{  \label{FigCurrentApproximation}
  Comparison between the exact result for the current and the perturbative expansion of Section \ref{edge-currents} (dashed line) in the middle of a $40\times40$-lattice at temperature $T/\gamma=0.1$ as a function of the distance $l$ from the sample edge. The main plot shows the current for a magnetic field strength of $\phi/\phi_0=10^{-3} \, \pi$, while the inset displays the results for $\phi/\phi_0=10^{-2} \, \pi$.}
\end{figure}

Even in the expansion of this order one can see many features imposed on the system by the magnetic field. A comparison between the full current and the approximation can be found in Fig.\ref{FigCurrentDecay} and Fig.~\ref{FigCurrentApproximation}. As expected, the current decays exponentially with the distance from the sample edge. It turns out, that the perturbative approach works surprisingly well yielding a good approximation for the edge current up to the depths of about 10\% of the sample size, see Fig.~\ref{FigCurrentDecay}. 

\section{Kitaev chain model}
\label{Kitaev_chain_model}

Recently Kitaev chain model moved in the focal point of research as it represents one of the simplest realizations of non-local Majorana edge states.\cite{Kitaev2001,QuantumComputingReview2008,Alicea2012} Its Hamiltonian is given by
\begin{eqnarray}      \label{Kitaevx}
H_\text{Kitaev} = H_\text{tb} +  \sum_{i=1}^{N-1} \Delta ( e^{i \phi} \, c_{i}^\dag c^\dag_{i+1} + 
e^{-i \phi} \, 
c_{i+1} c_{i} ) \, ,
\end{eqnarray} 
where $H_\text{tb}$ is defined in (\ref{H0x}). $\Delta$ is the gap parameter and $\phi$ is the superconducting phase. The action can be written in the form (\ref{funin1}) after the introduction of composite fields of the form
$c^\dag(i \omega_n) = (c_1^\dag(i \omega_n), c_1(i \omega_n), \dots, c_N^\dag(i \omega_n),c_N(i \omega_n))$. The action kernel is
\begin{eqnarray}        
\label{iter0}
{\cal A}_N = \left(
\begin{array}{cccccc}
{D}_0^{-1} & \Lambda & 0 & \ddots & 0 & 0 \\
\Lambda^\dag &  {D}_0^{-1} & 0 & \ddots  & 0 & 0 \\
0 & \Lambda^\dag &  {D}_0^{-1} & \ddots & 0 & 0 \\
\ddots  & \ddots  & \ddots  & \ddots  & \ddots & \ddots \\
0  & 0  & 0  & \ddots  & {D}_0^{-1} & \Lambda \\
0 & 0 & 0   & \ddots&  \Lambda^\dag &  {D}_0^{-1}
\end{array}
\right) \, ,
\end{eqnarray}
where the inter-site coupling matrix is
\begin{eqnarray}
 \Lambda = \left(
\begin{array}{cc}
- \gamma & - \Delta e^{i \phi} \\
  \Delta  e^{-i \phi} &  \gamma
\end{array}
\right) \, ,
\end{eqnarray}
and $D_0^{-1} = \mbox{diag}(i \omega_n - \epsilon, i \omega_n+\epsilon)$ is the action of a single detached site. The diagonal $2 \times 2$ Nambu blocks of the inverse of ${\cal A}_N$ yield the local GFs of a respective site, on its diagonal are the electron and hole GFs, while the anomalous GFs are on the off-diagonals: 
\begin{eqnarray}						\label{Dblock}
D  = \left(
\begin{array}{cc}
G & G^+ \\
G^- & \widetilde{G}
 \end{array}
\right) \, , 
\end{eqnarray}
where we define the local Matsubara GFs in Nambu representation as
\begin{eqnarray}
G(\tau) = - \langle T_\tau c_k(\tau) c_k^\dag (0) \rangle\, , \, \, \,
G^+(\tau) = - \langle T_\tau c^\dag_k(\tau) c_k^\dag (0) \rangle \, ,    
\nonumber \\ \nonumber 
\widetilde{G}(\tau) = - \langle T_\tau c^\dag _k(\tau) c_k(0) \rangle \, , \, \, \,
G^-(\tau) = - \langle T_\tau c_k(\tau) c_k (0) \rangle \, . 
\end{eqnarray}
The non-local GFs are defined in the same way with the pair of $k$ indices replaced by the indices of the sites between which the respective GF is considered. Let us denote by $\widetilde{D}_N$ the GF of the kind (\ref{Dblock}) at the {\it left} outermost site of a chain with length $M$. Then the following recurrence relation holds:
\begin{eqnarray}               \label{recKitaev0}
 \widetilde{D}_{N+1}^{-1} = D_0^{-1} - \Lambda \, \widetilde{D}_{N} \, \Lambda^\dag \, .
\end{eqnarray}
It can, of course, be recast into the Dyson equation as
\begin{eqnarray}        \label{DysonX}
 \widetilde{D}_{N+1} = D_0 + D_0 \, \Lambda \, \widetilde{D}_{N} \Lambda^\dag \, \widetilde{D}_{N+1} \, .
\end{eqnarray}
After making the substitution 
\begin{eqnarray}                       \label{SubsT}
\widetilde{D}_{N+1} = (\Lambda^\dag)^{-1} \widetilde{P}_{N} \widetilde{P}_{N+1}^{-1} \, ,
\end{eqnarray}
one then obtains a matrix-valued three-point recurrence relation for a new variable $\widetilde{P}_N$,
\begin{eqnarray}    \label{ChebyshevMatrix}
  \widetilde{P}_{N+1} &=& D_0^{-1} (\Lambda^\dag)^{-1}  \widetilde{P}_{N} - \Lambda (\Lambda^\dag)^{-1}  \widetilde{P}_{N-1}
 \nonumber \\
 &=& A \,  \widetilde{P}_{N} - B \,  \widetilde{P}_{N-1}
  \, . 
\end{eqnarray}
It can be considered to be a matrix generalization of the Chebyshev polynomials of the second kind. Using a similar approach we can compute the GF on the outmost {\it right} site of the chain. The corresponding identities are obtained by the exchange $\Lambda \leftrightarrow \Lambda^\dag$. We shall denote these GFs by $D_N$ and the respective Chebyshev polynomials by $P_n$. 

Now we compute the local GF at the site $1<k<N$ of the chain. Obviously, it cuts the chain into two pieces: (i) with length $k-1$ to the left and (ii) with length $N-k$ to the right of the site $k$. 
That is why $D_{k-1}$ and $\widetilde{D}_{N-k}$ are the corresponding self-energies and
\begin{eqnarray}							\label{arbitrary_kk}
 D_{kk} &=& \left( D_0^{-1} - \Lambda^\dag \, D_{k-1} \, \Lambda - \Lambda \, \widetilde{D}_{N-k} \, \Lambda^\dag \right)^{-1} 
   \\ \nonumber
 &=& 
  \left[ P_k \, P_{k-1}^{-1} \, \Lambda  - \Lambda \, (\Lambda^\dag)^{-1} \widetilde{P}_{N-k-1} \widetilde{P}_{N-k}^{-1} \, \Lambda^\dag \right]^{-1} \, .
 \, 
\end{eqnarray}
Using similar procedures one can derive an explicit expression for the GF between arbitrary sites $k$ and $m$. It is given by
\begin{widetext}
\begin{eqnarray}								\label{GkmResult}
 G_{km} &=& (-1)^{m-k} \, \left[ P_k \, P_{k-1}^{-1} \, \Lambda  - \Lambda \, (\Lambda^\dag)^{-1} \widetilde{P}_{N-k-1} \widetilde{P}_{N-k}^{-1} \, \Lambda^\dag \right]^{-1} \, \left(
P_0 \, P_{m-k-1}^{-1} \, \Lambda \right)
\nonumber \\
&\times&
 \left[ \widetilde{P}_{N-m+1} \, \widetilde{P}_{N-m}^{-1} \, \Lambda^\dag  - \Lambda^\dag \, \Lambda^{-1} P_{m-k-2} P_{m-k-1}^{-1} \, \Lambda \right]^{-1} \, .
\end{eqnarray}
\end{widetext}
For the detailed derivation see Appendix \ref{AppeC}. One special case: $k=1$, $m=N$ is particularly interesting as the respective GF is responsible for the transport properties of the chain. Here the expression is very appealing and concise: 
\begin{eqnarray}								\label{G1MResult}
 G_{1N} = (-1)^{N+1}  (\Lambda^\dag)^{-1} \, \widetilde{P}_{N-1} \,  \widetilde{P}_{N}^{-1}  \, P_{N-1}^{-1} \, . 
\end{eqnarray}
It is not difficult to show that in the scalar case the last two results immediately reproduce the corresponding formula for the tight-binding chain (\ref{throughU}). 
One of the applications for that is the derivation of the effective action for end Majoranas used in Refs.~[\onlinecite{GolubHorovitz2011,Weithofer2014}].

The most important advantage of these results is that the computation of the Chebyshev matrix polynomials itself only requires matrix multiplications. Only the very last steps in (\ref{GkmResult}) and (\ref{G1MResult}) require matrix inversions. This is different from directly using the recurrence relation (\ref{recKitaev}), which, being a generalization of a continuous fraction to matrices, requires a matrix inversion in each step. Needless to say, it is also more efficient than the direct matrix inversion of (\ref{iter0}). 

Just as in the case of the simple tight binding chain one can produce a bulk-boundary correspondence relation using the Dyson equation (\ref{DysonX}).
\begin{eqnarray}               \label{recKitaev}
 D_{\rm bulk}^{-1} = D_0^{-1} - \Lambda \, \widetilde{D}_{\rm edge} \, \Lambda^\dag - \Lambda^\dag \, D_{\rm edge} \, \Lambda \, ,
\end{eqnarray}
with the only difference that in the present situation there are two different edges coupled to a bulk site: a `right' and a `left' one.  

In order to access the end site GF we can alternatively follow the route discussed in the previous section and define new matrices $P_N$ and $Q_N$, similar to those in Eq.~(\ref{subs1}), so that $D_N = P_N Q_N^{-1}$. Then the recurrence relation (\ref{recKitaev0}) is solved by
\begin{eqnarray}
\label{matrixSol1Kitaev}
 \left( 
\begin{array}{c}
P_{N+1} \\
Q_{N+1}
\end{array}
\right) = R^N \,  \left( 
\begin{array}{c}
P_{1} \\
Q_{1}
\end{array}
\right) \, , \, \, \, \, \, 
\nonumber \\
R = \left(
\begin{array}{cc}
0 & ({\Lambda^\dag})^{-1} \\
- {\Lambda} & D_0^{-1} \, ({\Lambda^\dag})^{-1}
\end{array}
\right) \, ,
\end{eqnarray}
where $P_1 = D_0$ and $Q_1 = 1$. Further progress is made by diagonalizing the matrix $R$. Its eigenvalues are
\begin{widetext}
\begin{eqnarray}				\nonumber
 \lambda^2_{1,2} = 
 \frac{\pm [(i \omega_n)^2 - 2 (\gamma^2 + \Delta^2)] + 
 \sqrt{16 \Delta^2 \gamma^2 + (i \omega_n)^4 - 4 (i \omega_n)^2 (\gamma^2 + \Delta^2)}}{2 (\gamma^2 - \Delta^2)} \, .
\end{eqnarray}
\end{widetext}
Eigenvectors of $R$ can be written down as $({\bf u}, {\bf v})^T$. Obviously, 
\begin{eqnarray}    \label{ufromv}
{\bf u} = \lambda^{-1}   ({\Lambda^\dag})^{-1}�\, {\bf v} \, . 
\end{eqnarray}
From the requirement 
\begin{eqnarray}				\nonumber
 \left[
 D_0^{-1} \, ({\Lambda^\dag})^{-1} - \lambda \mathbb{1}
 - \lambda^{-1} {\Lambda} \, ({\Lambda^\dag})^{-1}
\right] {\bf v} = 0
\end{eqnarray}
we obtain 
\begin{eqnarray}				\nonumber
 v_2 = e^{- i \phi} \frac{(i \omega_n) \lambda \gamma - (1 + \lambda^2) \gamma^2 + (\lambda^2 - 1) \Delta^2}{ [(i \omega_n) \lambda - 2 \gamma] \Delta} v_1 \, .
\end{eqnarray}
Setting $v_1=1$ we obtain a set of 4 different vectors ${\bf v}(\lambda_{1,2,3,4})$, from which we compute ${\bf u}(\lambda_{1,2,3,4})$ using (\ref{ufromv}). Then 
\begin{eqnarray}				\nonumber
 T = \left(
 \begin{array}{cccc}
 {\bf u}(\lambda_{1}) &  {\bf u}(\lambda_{2}) &  {\bf u}(\lambda_{3}) &  {\bf u}(\lambda_{4}) \\
  {\bf v}(\lambda_{1}) &  {\bf v}(\lambda_{2}) &  {\bf v}(\lambda_{3}) &  {\bf v}(\lambda_{4})
\end{array}
 \right) \, , 
\end{eqnarray}
is the matrix which diagonalizes $R$. Thus we obtain 
\begin{eqnarray}				\nonumber
 R^N = T \, \mbox{diag} (\lambda_1^N, \lambda_2^N, \lambda_3^N, \lambda_4^N) \, T^{-1} \, .
\end{eqnarray}
In this way we obtain an analytical solution for the local GF in a Kitaev chain of {\it finite} length $N$.

\section{2D $p$-wave superconductor}
\label{pwave}

Kitaev chain model dealt with in the previous section is a 1D version of the more general $p$-wave superconductor models. In 2D it can be understood as a stack of Kitaev chains coupled by superconductor pairing (see e.~g. [\onlinecite{bernevig2013topological}]):
\begin{eqnarray}      \label{Hpwave}
 H_{\rm pw} &=& H_{\rm tb} 
+
   \sum_{n=1}^{N-1} \sum_{m=1}^{M-1}
 \left( i \Delta c_{n, m}^\dag c^\dag_{n, m+1} 
- i \Delta^*
 c_{n, m+1} c_{n, m}  \right.
\nonumber \\
&+& \left. 
\Delta c_{n, m}^\dag c^\dag_{n+1, m} 
+ \Delta^*
c_{n+1, m} c_{n, m}  \right)
 \, ,
\end{eqnarray} 
where $H_{\rm tb}$ is defined in Eq.~(\ref{H1x}). The system can be considered to be build up from $M$ Kitaev chains of length $N$, which are coupled by matrices $\boldsymbol{\Gamma}$:
\begin{eqnarray}
 \boldsymbol{\Gamma} = \mbox{diag}_N \left(
 \Lambda', \Lambda', \dots, \Lambda'
 \right) \, , \, \, \, \, \,
 \Lambda' = \left(
\begin{array}{cc}
 \gamma & i \Delta \\
 i \Delta^* & - \gamma
\end{array}
\right) \, .
\end{eqnarray}
Then the recurrence relation for the GF of the edge row is formally equivalent to that given in Eq.~(\ref{Dyson2}):
\begin{eqnarray}				 \label{Dyson_pwave}
{\bf g}_M^{-1} = {\bf G}^{-1} -  \boldsymbol{\Gamma}^\dag \, {\bf g}_{M-1} \,
  \boldsymbol{\Gamma}  \, ,
\end{eqnarray}
where ${\bf G}$ is the GF of an individual uncoupled Kitaev chain computed in (\ref{GkmResult}). The solution of this recurrence relation can again be performed using yet another set of Chebyshev  matrix polynomials of the second kind.

As an application we compute the energy spectra of systems of different size. As can be seen from 
\begin{figure}[h]
  \centering    \includegraphics[width=1. \columnwidth]{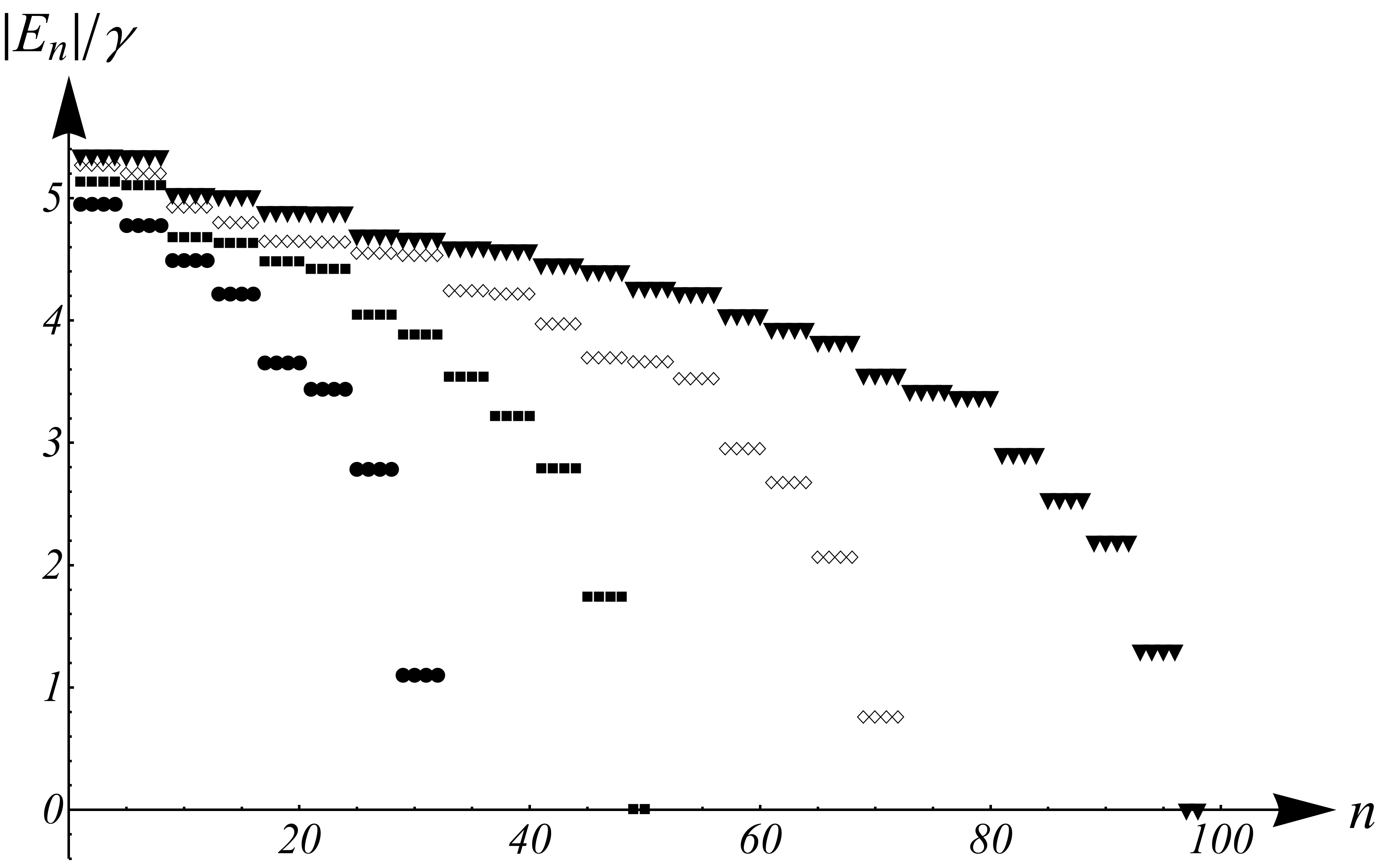}
  \caption{  \label{pWave_Energies}
  Energy levels of 2D $p$-wave superconductor systems for difference lattice sizes: $N=M=4,5,6,7$ (circles, squares, open diamonds and triangles) for $\mu=0$ and $\Delta/\gamma=2$. For $N=M=5,7$ there is a doubly degenerate energy level at $E=0$. 
  }
\end{figure}
Fig.~\ref{pWave_Energies} there is a pronounced even-odd effect. While for even $N=M$ there are no zero modes, for odd $N=M$ there is always a double-degenerate energy level at $E=0$. For more generic lattice sizes the zero modes exist whenever both $N$ and $M$ are odd. 
The case of $M=1$ corresponds to an ordinary Kitaev chain, in which the zero modes lie at precisely $E=0$ for odd $N$ and approach zero energy with growing even $N$.  A similar phenomenon takes place in the present case -- also in the case of either of $N$ or $M$ being even (or both) the eigenenergies tend towards $E=0$ with growing lattice sizes. There is, however, a fundamental difference between these lattices and those with both $N,M$ being odd. While the former are fourfold degenerate, the latter are always doubly degenerate. 
\begin{widetext}
The present 2D case is obviously different from the edge state in the Kitaev chain. It is instructive to investigate the spatial distribution of the DOS in order to find out whether the zero modes are localized. As we have shown above in Fig.~\ref{Fig1} there are three distinct points of the lattice: (a) corner site, (b) edge bulk site and (c) the true bulk site. In Fig.~\ref{DOS} we plot the corresponding DOS. Surprisingly, the zero energy states are indeed localized at the lattice corners and the respective DOS falls off exponentially with the distance from the corner.  We would like to stress that for large $N$ and $M$ there are always states in vicinity of $E=0$. But only the odd-odd constellation possesses a true zero mode, which is doubly degenerate. In all other cases there are multiple levels (at least four) in the vicinity of $E=0$.

\begin{figure}[h]
\centering

\begin{minipage}{0.45\textwidth}
\centering
\includegraphics[width=0.95\linewidth, height=0.2\textheight]{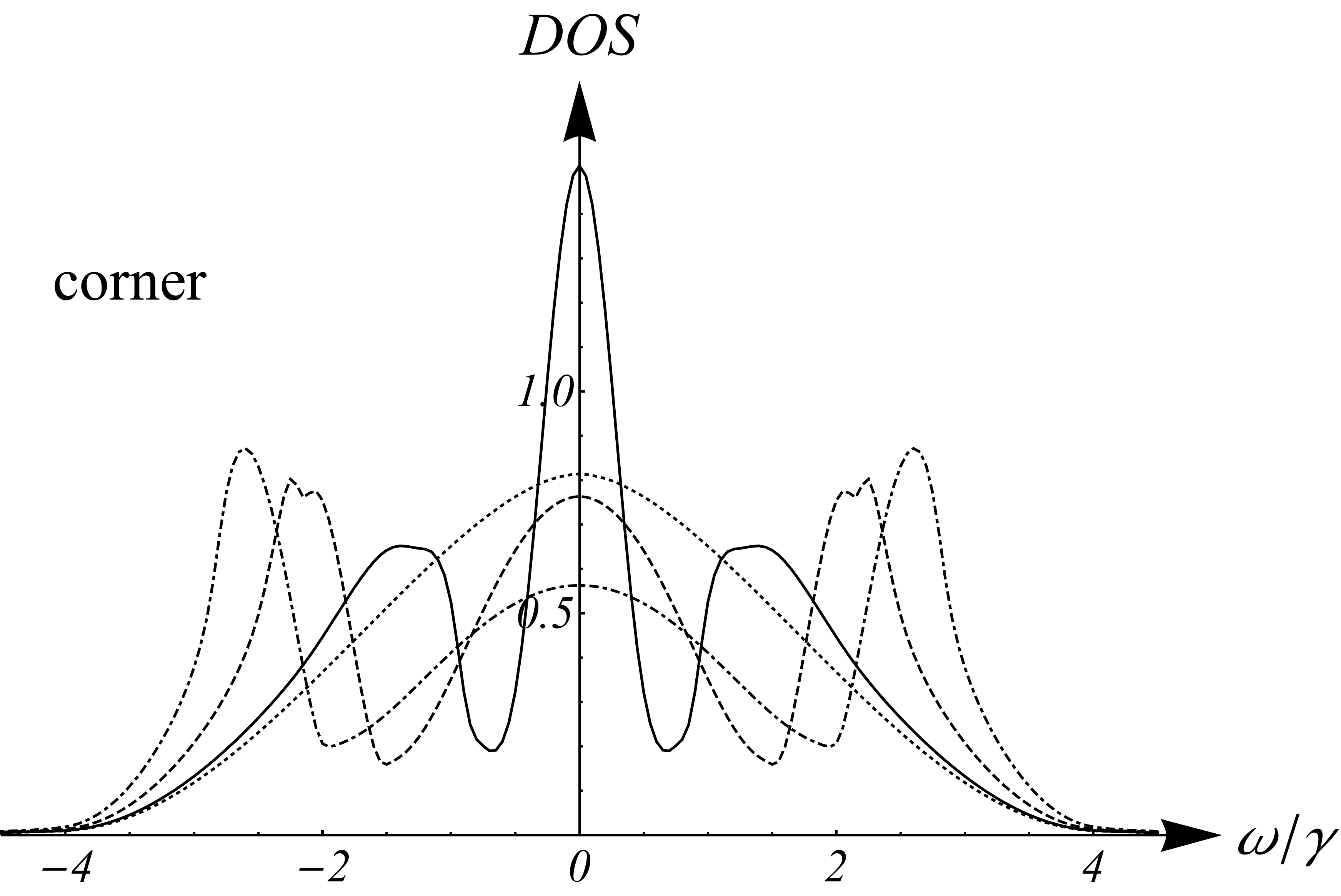}
\label{fig:prob1_6_2}
\end{minipage}
\begin{minipage}{0.45\textwidth}
\centering
\includegraphics[width=0.95\linewidth, height=0.2\textheight]{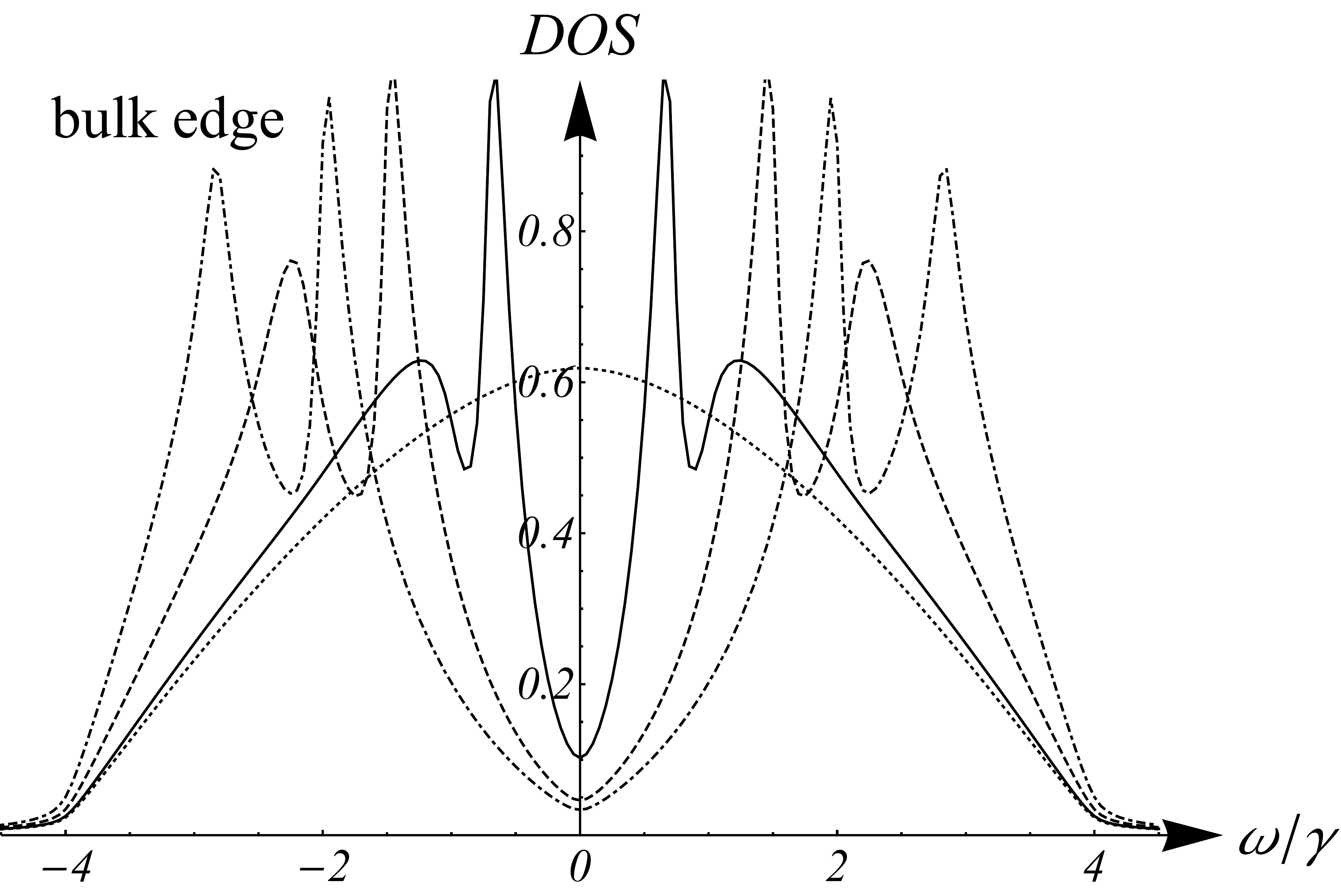}
\label{fig:prob1_6_1}
\end{minipage}

\begin{minipage}{0.45\textwidth}
\centering
\includegraphics[width=0.95\linewidth, height=0.2\textheight]{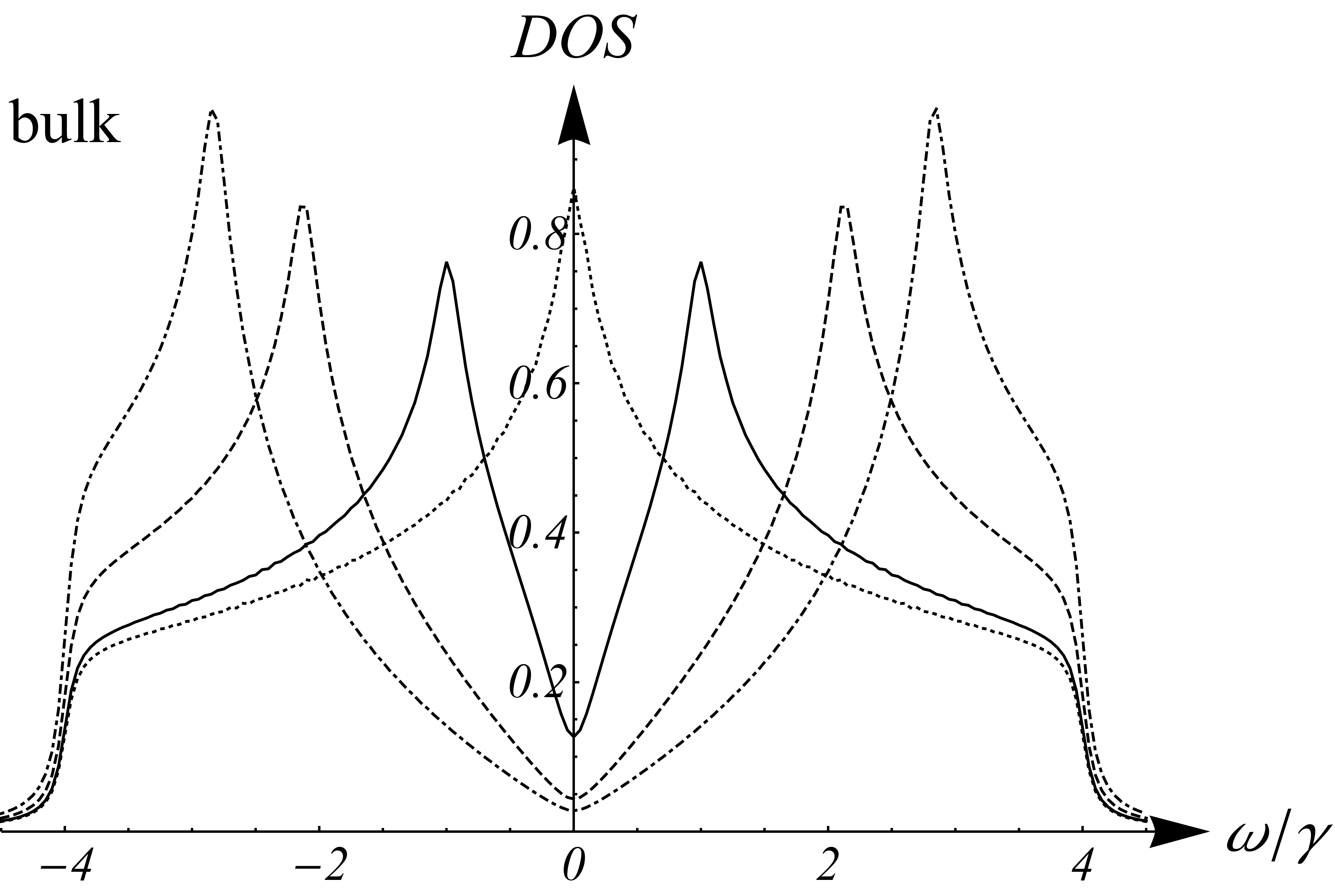}
\label{fig:prob1_6_2}
\end{minipage}
\begin{minipage}{0.45\textwidth}
\centering
\includegraphics[width=0.95\linewidth]{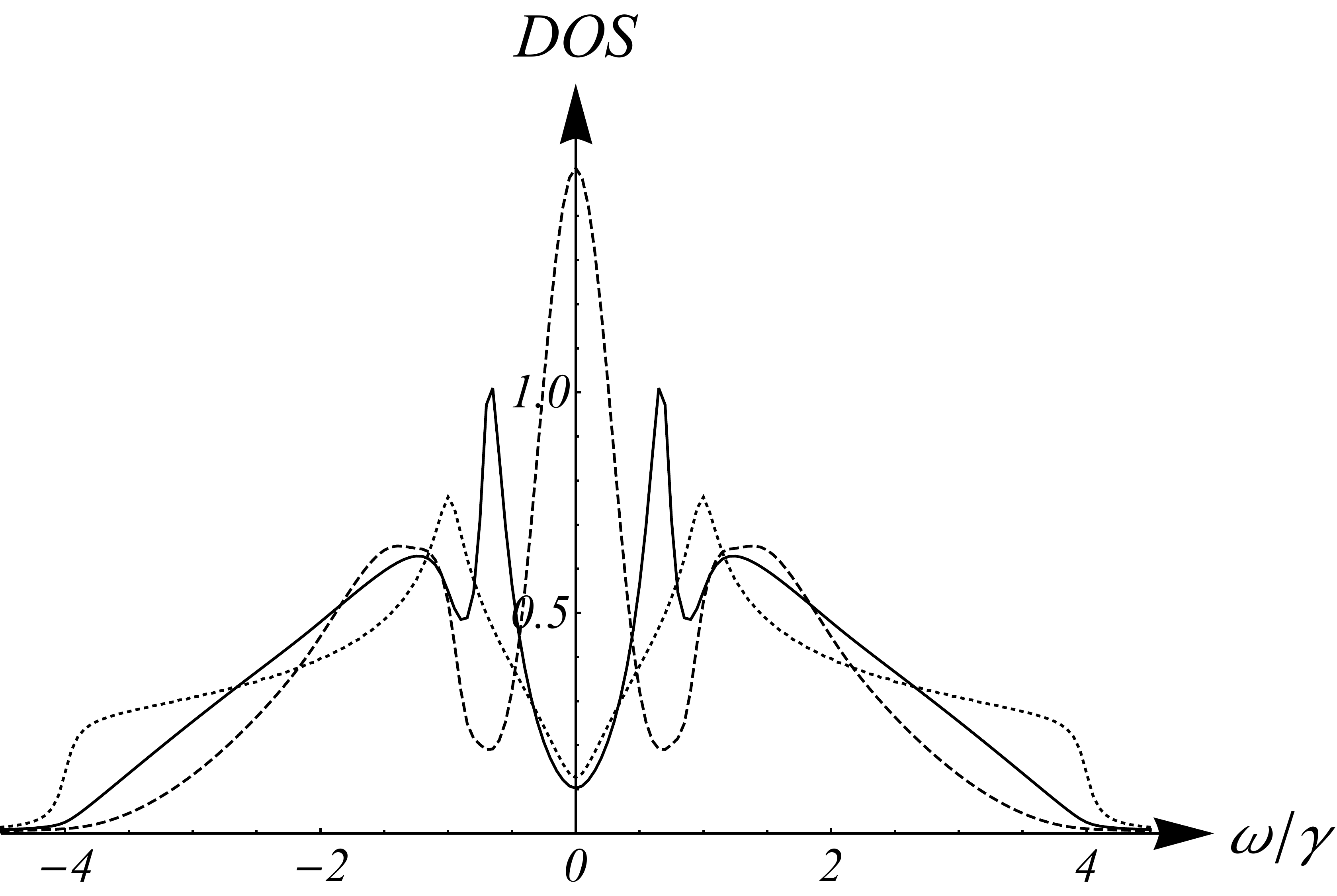}
\label{fig:prob1_6_1}
\end{minipage}
\caption{
\label{DOS}
DOS measured in arbitrary units as a function of energy at: (a) corner site, (b) bulk edge site, (c) bulk site in the center of the lattice for systems with dimensions $N=M=100$ and $\mu=0$, $\Delta/\gamma=0, 0.35, 0.75, 1$ (dotted, solid, dashed, dot-dashed lines, respectively). The lower right panel shows the comparison between the DOSes for $\Delta/\gamma=0.35$ at the corner, bulk edge and bulk sites (dashed, solid and dotted lines, respectively). All energy levels are artificially widened by $\delta/\gamma=0.075$ for better readability of the plots.}

\end{figure}
\end{widetext}
In the odd-odd case one is confronted with a fermionic state, which is highly delocalized between the four corners of the lattice. This is very similar to the end states in the open Kitaev chain.\cite{Kitaev2001} However, in the 2D case the corner states are not Majorana fermions. Nonetheless, they are as well perfectly suitable to be used as qubit states in the context of quantum information technology.

\section{3D tight binding lattice}
\label{3Dcase}

The recurrence relation method can be straightforwardly generalized to lattices of higher dimensions. Here we consider a 3D cubic tight binding sample with dimensions $M \times N \times K$. We recover all known results and generate a number of new ones.
The expression for the single-particle GF between the sites with coordinates $(j p q)$ and $(j' p' q')$ is given by
\begin{widetext}
\begin{eqnarray}  					\label{3DGF}
 g_{(j p q),(j' p' q')}(i \omega_n) 
&=& \sum_{s=1}^K
 \sum_{r=1}^N 
   \sum_{k=1}^M
 \frac{ v_{j k} v_{k j'} \, w_{p r} \, w_{r p'} \, y_{q s} \, y_{s q'}}{\frac{i \omega_n}{2} - \cos \left( \frac{\pi k}{M+1} \right) - \cos \left( \frac{\pi r}{N+1} \right)
- \cos \left( \frac{\pi s}{K+1} \right)}
\nonumber \\      			 \nonumber		&=&  \sum_{r=1}^N 
\sum_{k=1}^M v_{j k} v_{k j'} \, w_{p r} \, w_{r p'}
 \frac{ U_{q-1} \left( \epsilon_{k r} \right) \,  U_{K-q'} \left( \epsilon_{k r} 
  \right)}{  U_{K} \left( \epsilon_{k r} \right)} \, ,
\end{eqnarray}
\end{widetext}
where 
\begin{eqnarray}				\nonumber
y_{q s} =  \sqrt{\frac{2}{K+1}} \sin \left( \frac{\pi \, q \, s}{K+1} \right) \, ,
\end{eqnarray}
and
\begin{eqnarray}				\nonumber
 \epsilon_{k r} = \frac{i \omega_n}{2} - \cos \left( \frac{\pi k}{M+1} \right) - \cos \left( \frac{\pi r}{N+1} \right) \, .
\end{eqnarray}
For the DOS at the corner site of the lattice we then obtain the following result
\begin{eqnarray}									\label{CornerNu}
 \nu_{\rm corner}(\omega) &=&  4 \left(\frac{2}{\pi} \right)^2
\int \int d y d z \, \sqrt{(1-y^2)(1-z^2)}
\nonumber \\
&\times& \sqrt{1 - (\omega/2 - y - z)^2} \, , 
\end{eqnarray}
whereby the integration domain is fixed by the requirements $|y|<1$, $|z|<1$ and $|\omega/2 - y - z|<1$. The edge bulk site is the one on the edge of the sample far away from the corners 
\begin{eqnarray}									\label{eb3D}
\nu_{\rm eb}(\omega) &=& 2 \left(\frac{2}{\pi} \right)^2 \int \int  d y d z \sqrt{\frac{1-y^2}{1-z^2}}
\nonumber \\
&\times& 
\, \sqrt{1 - (\omega/2 - y - z)^2} \, , 
\end{eqnarray}
with the same integration domain. In the bulk of the face far away from the edges we obtain 
\begin{eqnarray}					\label{eb3D2}
\nu_{\rm face}(\omega) =  \left(\frac{2}{\pi} \right)^2  \int \int  d y d z 
\sqrt{\frac{1 - (\omega/2 - y - z)^2}{(1-y^2)(1-z^2)}} \, .
\end{eqnarray}
And, finally, in the bulk of the lattice one finds
\begin{eqnarray}
 \nu_{\rm bulk}(\omega) &=&  \frac{1}{2} \left(\frac{2}{\pi} \right)^2
\int \int d y d z \, \frac{1}{\sqrt{(1-y^2)(1-z^2)}}
\nonumber \\
&\times& 
\frac{1}{\sqrt{1 - (\omega/2 - y - z)^2}} \, .
\end{eqnarray}
To the best of our knowledge Eqs.~(\ref{CornerNu}) and (\ref{eb3D}) represent new results and are complementary to e.~g. those of  [\onlinecite{Guttmann2010}].
All remaining integrals can be rewritten in terms of elliptic integrals. We refrain from that though as it does not produce any added value.

\section{Conclusions}

We revisit the recurrence relation method (also referred to as transfer matrix method) for band structure calculations of lattice models and apply it for the computation of Green's functions (GF). We show a number of analytical solutions for conventional lattices in different dimensions. While for a simple 1D tight-binding system every single GF can be written down as a rational function of Chebyshev polynomials of the second kind, in higher dimensions or for systems with such non-trivial structure as a $p$-wave superconductor or ones with spin-orbit coupling the resulting expressions for the GFs are given by functions of matrix-valued Chebyshev polynomials. Using this concept we derive an explicit expression for any kind of GF for a Kitaev chain, see Eq.~(\ref{GkmResult}). Even though our results require numerical calculations, their efficiency is vastly superior to all existing methods as they require only small number of matrix inversions, whatever the system size. 

We generalize the method for lattices in external fields and for corresponding GFs derive analytical formulas in closed form. By an explicit computation of the density of states and energy-resolved particle currents we show how in a 2D tight binding lattice subject to a magnetic field a 1D chiral edge state is formed. Furthermore, we develop a perturbative approach in order to construct simpler analytical solutions, which adequately describe the properties of edge states for not too strong fields. 

Application of our method to a 2D $p$-wave superconductor on a lattice reveals its very interesting energy level structure. It turns out, that for lattices with odd length and width there is always a doubly degenerate zero energy state, which is non-local and spread between the four corners of the lattice. That is confirmed by an explicit calculation of the spatial dependence of the respective density of states. This phenomenon is very similar to end states in open Kitaev chains. However, the emergent fermionic state is not of Majorana type. Nonetheless, its non-locality is very advantageous for future quantum information technology applications, not least due to its explicit higher dimensionality, which entails a better experimental feasibility.

In addition we derive numerous analytical results for densities of states of conventional tight-binding lattices in different dimensions and different spatial locations in lattices with open boundaries.

\acknowledgements
AK is supported by the Heisenberg Programme of the Deutsche Forschungsgemeinschaft (Germany) under Grant No. KO 2235/5-1.

\appendix 
\section{}
\label{AppB}

With the periodic boundary condition the action matrix is slightly modified: 
\begin{eqnarray}					\label{Amatrix1}				\nonumber
{\cal A}  = \left(
\begin{array}{cccccc}
i \omega_n & - \gamma & 0 & \ddots & 0 & - \gamma \\
- \gamma & i \omega_n & - \gamma & \ddots & 0 & 0 \\
0 & - \gamma & i \omega_n & \ddots &0 & 0 \\
\ddots & \ddots & \ddots & \ddots & \ddots & \ddots \\
0 & 0 & 0 & \ddots & i \omega_n &  - \gamma \\
-\gamma & 0 & 0 & \ddots & - \gamma & i \omega_n 
\end{array} \right) \, .
\end{eqnarray}
Its determinant and the inverse is given by
\begin{eqnarray}				\nonumber
\mbox{det} \, {\cal A} = 2 U_{M}\left(\frac{i \omega_n}{2\gamma}\right) - \left(\frac{i \omega_n}{\gamma}\right)U_{M-1}\left(\frac{i \omega_n}{2\gamma}\right) - 2 \, ,
 \\ \nonumber
 ({\cal A}^{-1})_{km} = \frac{ U_{N - 1 - |k-m|}\left(\frac{i \omega_n}{2\gamma}\right) +  U_{|k-m|-1}\left(\frac{i \omega_n}{2\gamma}\right)}{2 U_{N}\left(\frac{i \omega_n}{2\gamma}\right) - \left(\frac{i \omega_n}{\gamma}\right) U_{N-1}\left(\frac{i \omega_n}{2\gamma}\right) - 2} \, .
\end{eqnarray}
On the other hand the matrix 
\begin{eqnarray}				\nonumber
  S_{k m} = e^{i 2 \pi k m /N} \, 
\end{eqnarray}
straightforwardly diagonalizes ${\cal A}$. That is why the GF is also given by the following expression:
\begin{eqnarray}  					\label{periodic1D}
 g_{km}(i \omega_n) = \sum_{l=1}^N S^{-1}_{kl} \frac{1}{i \omega_n  - \epsilon_l} S_{lm}
 = \sum_{l=1}^N \frac{e^{ i 2 \pi l (k - m)/N}}{i \omega_n  - \epsilon_l} \, ,
\end{eqnarray}
where $\epsilon_k = - 2 \gamma \cos( 2 \pi k /N)$. From the equality $ g_{km}(i \omega_n) = ({\cal A}^{-1})_{km}$ then follows an interesting result for the trigonometric sum in (\ref{periodic1D}).

\section{}   
\label{AppeA}

The retarded GF obtained from the Matsubara GF in (\ref{DOSedge1D}) can be found to be
\begin{eqnarray}
g^R(\omega) = \frac{1}{\gamma^2} \left\{
  \begin{array}{cl}
   \left[ \frac{\omega}{2}  - i \sqrt{\gamma^2 - \left(  \frac{\omega}{2} \right)^2}  \right]
   & |\omega|<2 \gamma \\
     \left[ \frac{\omega}{2}  + \sqrt{\left( \frac{\omega}{2} \right)^2 - \gamma^2}  \right]
 & \omega< -2 \gamma \\
   \left[ \frac{\omega}{2}  - \sqrt{\left( \frac{\omega}{2} \right)^2 - \gamma^2} \right]
 & \omega> 2 \gamma \\
  \end{array}
  \right.
\end{eqnarray}
As a result the DOS is only nonzero within the band $|\omega|<2 \gamma$ and has the expected half-ellipsoidal form 
\begin{eqnarray}
\nu_{\rm edge}(\omega) = - 2 \, \mbox{Im} \, g^R(\omega) = \gamma^{-1} \sqrt{1 - \left(  \omega/2\gamma \right)^2}
\end{eqnarray}
In the bulk DOS one immediately recognizes the van Hove singularities, 
\begin{equation}							\label{1DlocalDosBulk}
\nu_{\rm bulk}(\omega) = \left[ \gamma \, \sqrt{1 - \left( \omega / 2\gamma \right)^2} \right]^{-1} \, .
\end{equation}

\begin{widetext}
\section{}   
\label{AppeC}

We first outline the calculation of the end-to-end GF. Here we use the chain contraction procedure proposed in \cite{MajoranaAK2016}. One starts with a partition function generated by the action (\ref{iter0}) and subsequently integrates out all fermionic fields up to those describing the end sites. We start with
\begin{eqnarray}        \label{iter3}				\nonumber
{\cal A}_{M} = \left(
\begin{array}{cccccc}
{D}_0^{-1} & T_1 & 0 & \ddots & 0 &0 \\
\widetilde{T}_1 &  {F}_1^{-1} & \Lambda & \ddots  & 0 & 0 \\
0 & \Lambda^\dag &  {D}_0^{-1} & \ddots  & 0 & 0 \\
\ddots& \ddots& \ddots&  \ddots & \ddots & \ddots\\
0  & 0  & 0  & \ddots  & {D}_0^{-1} & \Lambda \\
  0 & 0 & 0   & \ddots&  \Lambda^\dag &  {D}_0^{-1}
\end{array}
\right) \, .
\end{eqnarray}
Then, after integrating out the fermions of the second site (counted from the left) the action kernel is given by a matrix of reduced dimensions:
\begin{eqnarray}        \label{iter4}				\nonumber
{\cal A}_{M-1} &=& \left(
\begin{array}{cccccc}
{D}_0^{-1} - T_1 F_1 \widetilde{T}_1 & -T_1 F_1 \Lambda & 0 & \ddots & 0 &  0 \\
-\Lambda^\dag F_1 \widetilde{T}_1  & D_0^{-1} - \Lambda^\dag {F}_1 \Lambda & \Lambda & \ddots  & 0 & 0 \\
0 & \Lambda^\dag &  {D}_0^{-1} & \ddots  & 0 & 0 \\
\ddots  & \ddots  & \ddots  &  \ddots & \ddots  & \ddots  \\
0  & 0  & 0  & \ddots  & {D}_0^{-1} & \Lambda \\
 0 & 0 & 0   & \ddots&  \Lambda^\dag &  {D}_0^{-1}
\end{array}
\right) 
\nonumber \\
&=& \left(
\begin{array}{cccccc}
{D}_0^{-1} - T_1 F_1 \widetilde{T}_1 & T_2 & 0 & \ddots & 0 & 0 \\
\widetilde{T}_2 &  {F}_2^{-1} & \Lambda & \ddots & 0 & 0 \\
0 & \Lambda^\dag &  {D}_0^{-1} & \ddots & 0 & 0 \\
\ddots & \ddots & \ddots &  \ddots  & \ddots & \ddots \\
0  & 0  & 0  & \ddots  & {D}_0^{-1}  & \Lambda \\
  0 & 0 & 0   & \ddots&  \Lambda^\dag &  {D}_0^{-1}
\end{array}
\right)
\, .
\end{eqnarray}
Continuing this lattice contraction we finally obtain the $2 \times 2$ action
\begin{eqnarray}					\label{A2}
   {\cal A}_{2} &=& \left(
\begin{array}{cc}
 D_0^{-1} - \sum\limits_{j=1}^{N-2} T_j \, F_j \, \widetilde{T}_j & - T_{N-2} \, F_{N-2} \, \Lambda \\
 -  \Lambda^\dag \, F_{N-2} \, \widetilde{T}_{N-2} & D_0^{-1} - \Lambda^\dag \, F_{N-2} \, \Lambda
\end{array}
\right) \, , 								\nonumber
\end{eqnarray}
where for $F_n$, $T_n$ and $\widetilde{T}_n$ we have the following recurrence relations: 
\begin{eqnarray}						\label{RR0}
 F_{n+1}^{-1} &=& D_0^{-1} - \Lambda^\dag \, F_n \, \Lambda \, , \, \, \, \, \, 
 \nonumber \\
 T_{n+1} &=& - T_n \, F_n \, \Lambda \, , \, \, \, \, \, 
 \nonumber \\
  \widetilde{T}_{n+1} &=& - \Lambda^\dag \, F_n \, \widetilde{T}_n \, 		
\end{eqnarray}
with the initial conditions $F_1=D_0$, $T_1 = \Lambda$ and $\widetilde{T}_1 = \Lambda^\dag$. So $F_n$ satisfy the recurrence relation for the local GF on the outmost right site $D_n$, see Section \ref{Kitaev_chain_model}. Now we use the formula for the block matrix inversion
\begin{eqnarray} 						\label{matinversion}				\nonumber
 \left(
\begin{array}{cc}
  A & B \\
  C & D
\end{array}
\right)^{-1} 
= 
\left(
\begin{array}{cc}
 (A - B D^{-1} C )^{-1} & - (A - B D^{-1} C )^{-1} B D^{-1} \\
 - D^{-1} C (A - B D^{-1} C )^{-1} & D^{-1} + D^{-1} C (A - B D^{-1} C )^{-1} B D^{-1}
\end{array}
\right) \, ,
\end{eqnarray}
where $A, B, C, D$ are arbitrary non-singular quadratic matrices. Applying this to invert (\ref{A2})
we immediately recognize, that 
\begin{eqnarray}
  (A - B D^{-1} C )^{-1} \equiv \widetilde{D}_N \, ,
\end{eqnarray}
i. e. it is equal to the local GF of the outmost left site of the chain. That is why for the object $G_{1N}$ we obtain the relation
\begin{eqnarray}				\nonumber
 G_{1N} = - (A - B D^{-1} C )^{-1} B D^{-1} =
 - \widetilde{D}_N \, B \, D_{N-1} 
 \nonumber \\
 =  (-1)^{N+1}   (\Lambda^\dag)^{-1} \, \widetilde{P}_{N-1} \,  \widetilde{P}_{N}^{-1}  \, P_{N-1}^{-1} \, .
\end{eqnarray}

In order to access the $k$-to-$m$ GF for $m>k$ we follow the following strategy. 
First one integrates out all sites with indices $j<k$ and $j>m$. This process can be understood as {\it outer} chain contractions. Then one is confronted with the effective action
\begin{eqnarray}				\nonumber
{\cal A}_{km} = \left(
\begin{array}{ccccc}
{D}_{k}^{-1} & \Lambda & \ddots & 0 & 0 \\ 
\Lambda^\dag &  {D}_0^{-1} &  \ddots & 0 & 0\\
\ddots & \ddots & \ddots & \ddots & \ddots\\
0& 0 & \ddots & {D}_0^{-1} & \Lambda \\
0 & 0 & \ddots & \Lambda^\dag &  \widetilde{D}_{N-m+1}^{-1}  \\
\end{array}
\right) \, .
\end{eqnarray} 
Our task is now the calculation of the end-to-end GF for this action. To this end we employ the approach used above, which we call {\it inner} contractions. As a result, for the end-to-end effective action kernel we get
\begin{eqnarray}				\nonumber
 \label{A2km}
   {\cal A}_{2} &=& \left(
\begin{array}{cc}
 D_k^{-1} - \sum\limits_{j=1}^{m-k-1} T_j \, F_j \, \widetilde{T}_j & - T_{m-k-1} \, F_{m-k-1} \, \Lambda \\
 -  \Lambda^\dag \, F_{m-k-1} \, \widetilde{T}_{m-k-1} & \widetilde{D}_{N-m+1}^{-1} - \Lambda^\dag \, F_{m-k-1} \, \Lambda
\end{array}
\right) \, .  
\end{eqnarray}
Inversion of this expression yields for the off-diagonal component the value
\begin{eqnarray}				\nonumber
 G_{km} &=& (-1)^{m-k} \, \left[ P_k \, P_{k-1}^{-1} \, \Lambda  - \Lambda \, (\Lambda^\dag)^{-1} \widetilde{P}_{N-k-1} \widetilde{P}_{N-k}^{-1} \, \Lambda^\dag \right]^{-1} \, \left(
P_0 \, P_{m-k-1}^{-1} \, \Lambda \right)
\nonumber \\
&\times&
 \left[ \widetilde{P}_{N-m+1} \, \widetilde{P}_{N-m}^{-1} \, \Lambda^\dag  - \Lambda^\dag \, \Lambda^{-1} P_{m-k-2} P_{m-k-1}^{-1} \, \Lambda \right]^{-1} \, .
\end{eqnarray}
We would like to remark that this kind of procedure can be used for inversion (and, of course, calculation of the determinants) of block-tridiagonal matrices even in the general case of arbitrary $\Lambda$, $\Lambda^\dag$ and $D_0$ matrices. Our approach is different from that presented in e.~g. [\onlinecite{Molinari97}].
\end{widetext}

\section*{References}

\bibliographystyle{revtex}
\bibliography{Lattice_fermions.bib}

\end{document}